\documentclass[10pt,twocolumn,letterpaper]{article}

\usepackage[pagenumbers]{cvpr}

\newcommand{\PAR}[1]{\vskip4pt \noindent{\bf #1~}}

\usepackage{colortbl}
\usepackage{float}
\definecolor{rkfirst}{RGB}{255,179,179}
\definecolor{rksecond}{RGB}{255,217,179}
\definecolor{rkthird}{RGB}{255,247,179}
\newcommand{\fst}[1]{\cellcolor{rkfirst}#1}
\newcommand{\snd}[1]{\cellcolor{rksecond}#1}
\newcommand{\trd}[1]{\cellcolor{rkthird}#1}
\usepackage{tikz}
\newcommand{\dashcell}[2]{\tikz[baseline=(X.base)]{\node[draw=#1,dash pattern=on 1.2pt off 0.8pt,line width=0.9pt,inner sep=0.6pt,outer sep=0pt](X){#2};}}
\newcommand{\dfst}[1]{\dashcell{rkfirst}{#1}}
\newcommand{\dsnd}[1]{\dashcell{rksecond}{#1}}
\newcommand{\dtrd}[1]{\dashcell{rkthird}{#1}}
\newcommand{\rfmark}{\ensuremath{^\ast}}
\newcommand{\pap}[1]{\textcolor{gray}{#1}}

\definecolor{cvprblue}{rgb}{0.21,0.49,0.74}
\usepackage[pagebackref,breaklinks,colorlinks,allcolors=cvprblue]{hyperref}

\def\confName{CVPR}
\def\confYear{2027}

\title{SimpleTet - A Simple Tetrahedral Mesh Representation for Novel View Synthesis}

\author{
Patrik Beliansky$^{1,2}$ \qquad Yaqing Ding$^{3}$ \qquad Zuzana Kukelova$^{4}$ \qquad Torsten Sattler$^{1}$ \\
$^{1}$ Czech Institute of Informatics, Robotics and Cybernetics, Czech Technical University in Prague \\
$^{2}$ Faculty of Electrical Engineering, Czech Technical University in Prague \\
$^{3}$ Southeast University \\
$^{4}$ Visual Recognition Group, Faculty of Electrical Engineering, Czech Technical University in Prague
}

\begin{document}
\maketitle
\begin{abstract}
Compared to 3D Gaussians, tetrahedral meshes enable potentially simpler and, due to exact visibility rendering, more efficient rendering pipelines.
In the context of novel view synthesis, optimizing representations based on tetrahedral meshes is more complicated compared to optimizing 3D Gaussians: Moving vertices can lead to intersections between the tetrahedra, while retriangulation can flip edges in the meshes, leading to discontinuities during the optimization. Handling such discontinuities is thus one focus of existing works, which design sophisticated approaches to handle these issues, \eg, using dual or additional volumetric representations. In contrast,
we propose SimpleTet, a conceptually simpler representation:
Densities and features are stored at the vertices, which move freely, and the mesh is retriangulated as they move.
We assume constant densities per tetrahedron, obtained as the
geometric mean of the densities of its vertices.
During rendering, we interpolate features barycentrically, and aggregate them along rays.
A small view-conditioned MLP is then used once for each pixel to decode the pixel's feature into a color.
On Mip-NeRF~360 and Tanks and Temples, our method reaches state-of-the-art quality among volumetric methods, while training efficiently.
We thus show that (1) a simple tetrahedral mesh with values at its vertices can perform as well as more complicated approaches, contrary to claims in the literature, and (2) that changing mesh topology during optimization does not need to be a significant problem.
\end{abstract}

\section{Introduction}
\label{sec:intro}

Novel-view synthesis aims to render a scene from new viewpoints given a set of input images.
State-of-the-art methods first reconstruct the scene from the given set of images (and their known poses).
NEural Radiance Fields (NeRFs)~\cite{mildenhall2020nerf,mueller2022instant}, 3D Gaussian splatting (3DGS)~\cite{kerbl3Dgaussians}, and their many variants, have popularized this reconstruction-based approach,
showing (close to) photorealistic novel-view synthesis of scenes
while efficiently learning their underlying scene representations from images.

In terms of scene representations, volumetric meshes, \eg, obtained by Delaunay triangulation, partition the scene space into non-overlapping cells~\cite{govindarajan2025radfoam,mai2026radiance,kulhanek2023tetranerf}.
Colors and densities can be integrated exactly over the segment of a ray that passes through such a  cell~\cite{govindarajan2025radfoam,mai2026radiance}.
This in turn enables volume rendering via rasterization~\cite{mai2026radiance}.
In the context of tetrahedral meshes, this means rendering triangles, which can be done very efficiently on GPUs.

Tetra-NeRF~\cite{kulhanek2023tetranerf} renders a static volumetric mesh, the Delaunay tetrahedralization of a sparse or dense set of 3D points, and stores features at its vertices.
Since the mesh is fixed, the quality of the reconstruction depends on the input points, and regions without points cannot be refined.
Recent methods therefore move and add the points during the optimization~\cite{govindarajan2025radfoam,govindarajan2026powerfoam,taveira2026vorotracing,mai2026radiance}.
These methods represent the scene using the Delaunay tetrahedralization of the current point set or its dual.
As the points move, local flips can change the tetrahedral connectivity, causing the volumetric field to change discontinuously.
Radiant Foam~\cite{govindarajan2025radfoam} and VoroTracing~\cite{taveira2026vorotracing} thus render Voronoi cells, while PowerFoam~\cite{govindarajan2026powerfoam} uses cells of a power diagram, whose shapes vary continuously with the positions of their sites.
Radiance Meshes~\cite{mai2026radiance} keeps the tetrahedra, but queries an instant-NGP field~\cite{mueller2022instant} for their values during training and bakes the values into the tetrahedra at the end, when the positions and tetrahedra are fixed.
These designs aim to avoid or reduce the impact discontinuities, and come at the cost of a different cell type or an additional field.

We show that \emph{SimpleTet}, a simpler tetrahedral mesh representation for novel view synthesis, matches or exceeds the rendering quality of these methods. As in Tetra-NeRF, each vertex stores a density and a feature vector. Following Radiance Meshes~\cite{mai2026radiance}, we retriangulate the mesh as new vertices are added or existing vertices move, and rasterize the tetrahedra in power-sorted order.
We assume a constant density per tetrahedron, defined as the geometric mean of the densities of its vertices.
In the later stages of the training process, we drop per-vertex densities in favor of optimizing per-tetrahedron densities directly.
We render the representation by integrating features along each ray and decoded the accumulated feature once per pixel using a small MLP.
New vertices are added where the rendering error is high, as in error-based densification of Gaussians~\cite{rotabulo2024revising,mallick2024taming}.

We make the following contributions
(\textbf{1}) We introduce SimpleTet, a tetrahedral mesh representation for novel view synthesis that jointly optimizes vertex positions, densities, and features. Densities and features remain attached to their vertices through retriangulation.
While~\cite{mai2026radiance} report that such a representation is inferior to more complicated approaches, SimpleTet reaches state-of-the-art performance. SimpleTet does not explicitly handle discontinuities, yet we do not observe significant issues during training.
(\textbf{2}) We compute each tetrahedron's density as the geometric mean of its vertex densities and use a small MLP to decode ray-aggregated features once per pixel. We ablate both design choices, showing that both significantly affect performance.
(\textbf{3}) We compare SimpleTet with four recent volumetric methods across all 30 scenes from Mip-NeRF~360 and Tanks and Temples. SimpleTet achieves the best average PSNR, SSIM, and LPIPS, with a training time close to that of the fastest baseline.
(\textbf{4}) We will release our full implementation, including our CUDA rasterizer.

\section{Related work}
\label{sec:related}

\noindent \textbf{Radiance fields.}
NeRF~\cite{mildenhall2020nerf} represents a scene as a neural field and renders it by sampling points along each ray.
Instant-NGP~\cite{mueller2022instant} replaces most of the network with a multiresolution hash grid.
3DGS~\cite{kerbl3Dgaussians} replaces the field with explicit Gaussians, which are rasterized.
3DGRT~\cite{moenne2024gaussian} ray traces the Gaussians, and EVER~\cite{mai2025ever} renders ellipsoids of constant density exactly along each ray.
3D Convex Splatting~\cite{held2025convex} and Triangle Splatting~\cite{held2025triangle} replace the Gaussians by smooth convex shapes and by triangles.
The methods below keep volume rendering, but split space into cells.

\noindent \textbf{Volumetric meshes.}
Tetra-NeRF~\cite{kulhanek2023tetranerf} builds a Delaunay tetrahedralization of from SfM points and keeps it fixed during training.
The vertices store features, which are interpolated barycentrically inside each tetrahedron.
It traces rays through the tetrahedra, samples points along them and decodes each sample with an MLP.
Radiance Meshes~\cite{mai2026radiance} optimizes the vertex positions and retriangulates the vertices during training.
Each tetrahedron has a constant density and a linear color, predicted by an instant-NGP hash grid and baked into per-tetrahedron values late in training.
Sorting the tetrahedra by power distance gives the exact visibility order, so they are rasterized instead of ray traced.

Our proposed SimpleTet approach combines the per-vertex features of Tetra-NeRF with  the moving vertices and the rasterization of Radiance Meshes, without using the hash grid.
With the values at the vertices, each tetrahedron obtains a density from the densities of its vertices.
MeshSplatting~\cite{Held2025MeshSplatting}, which renders opaque triangles, gives each triangle the minimum opacity of its vertices. We use the geometric mean instead to obtain per-tetrahedron densities as this significantly improves performance.

\cite{mai2026radiance} explicitly note that their approach outperforms a baseline that uses per-vertex features and densities.
In contrast, we show that such a baseline can perform similar to or better than their approach, while training faster.

\noindent \textbf{Voronoi cells.}
Radiant Foam~\cite{govindarajan2025radfoam} represents the scene as a Voronoi diagram of optimized points, each cell with a density and an Spherical Harmonics (SH) color, and traces rays through the cells.
PowerFoam~\cite{govindarajan2026powerfoam} uses a power diagram of oriented, textured cells, and uses both rasterization and ray tracing.
VoroTracing~\cite{taveira2026vorotracing}  traces rays through
Voronoi cells with octahedral appearance textures, initialized from dense correspondences, without densification or pruning.

These cell-based approaches are explicitly motivated by the discontinuities observed when maintaining a Delaunay triangulation during training~\cite{govindarajan2025radfoam}.
However, we do not observe significant issues during training despite not explicitly accounting for discontinuities.

\noindent \textbf{Densification.}
3DGS~\cite{kerbl3Dgaussians} clones and splits Gaussians with large view-space position gradients, and later work refines this gradient criterion~\cite{ye2024absgs,zhang2024pixelgs}.
Revising Densification~\cite{rotabulo2024revising} instead distributes a per-pixel SSIM error to the Gaussians by their blending weights and grows those above an error threshold, up to a budget.
Taming 3DGS~\cite{mallick2024taming} grows the model to a given budget by sampling Gaussians in proportion to a score, and 3DGS-MCMC~\cite{kheradmand2024mcmc} keeps a fixed budget and relocates Gaussians.
The cell-based methods densify in a similar way.
Radiant Foam samples the cells to split by the gradient norm of their positions, and PowerFoam resamples cells by their blending-weighted color error towards a growing target count.
Radiance Meshes adds vertices to the tetrahedra whose image error is above a threshold, up to a budget, between the rays of the two views with the most error.
We distribute the error by the blending weights as Revising Densification does, but instead of a fixed error threshold, our candidates are the tetrahedra with an above-average error in some view.
Like the above methods, we grow up to a budget, sample the candidates in proportion to their error as Taming 3DGS and PowerFoam do, and insert each vertex at the centroid of its tetrahedron.

\section{Method}
\label{sec:method}

\begin{figure*}[t]
\centering
\includegraphics[width=0.8\linewidth]{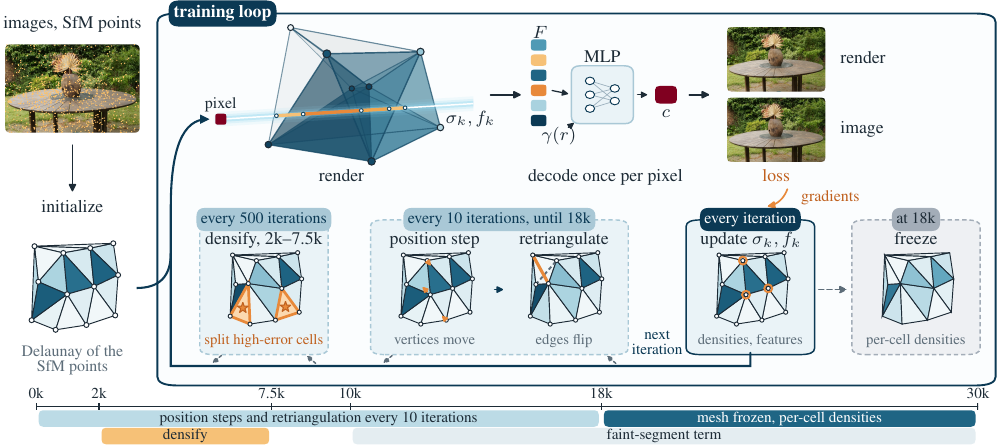}
\caption{\textbf{Overview of SimpleTet.}
Top, one training iteration renders each pixel by accumulating vertex features along its ray through the tetrahedra and decoding them once with an MLP.
Bottom, the updates of densities, features, and vertex positions after computing the loss and how often they run.
The timeline shows when each step is active.}
\label{fig:overview}
\end{figure*}

\Cref{fig:overview} illustrates SimpleTet.
We represent the scene as a volumetric mesh obtained by Delaunay triangulation~\cite{delaunay_alg_geometry}.
Following~\cite{kulhanek2023tetranerf}, each vertex in the mesh is associated with a density and a feature vector.
The density associated with a tetrahedron is constant and defined as the geometric mean of the densities stored at its vertices.
The feature vector for a 3D point inside of a tetrahedron is obtained via barycentric interpolation~\cite{kulhanek2023tetranerf}.
During rendering, rather than aggregating colors along rays, the features encountered along a ray are aggregated into a single feature vector for the corresponding pixel~\cite{thies2019deferred, hedman2021baking, reiser2023merf, chen2023mobilenerf, zhang2024ref}.
A small MLP then decodes the resulting feature to obtain the pixel's color.
Unlike~\cite{kulhanek2023tetranerf}, we allow the vertices to move during optimization.
We further add additional vertices to refine the mesh in regions that lead to large rendering errors. Compared to~\cite{mai2026radiance,govindarajan2025radfoam}, we do not devise any new strategies for treating potential discontinuities that arise when vertices are moved or inserted.
We show that when storing values at the vertices, simply retriangulating the scene and continuing training already leads to state-of-the-art results.

\subsection{Representation}
\label{sec:representation}
By construction, a Delaunay tetrahedralization of a point set has no point inside the circumsphere of any tetrahedron~\cite{delaunay_alg_geometry}.
For points in general position, it is unique, so the positions alone define the mesh.
Following prior work~\cite{kulhanek2023tetranerf,mai2026radiance}, we initialize our volumetric mesh with the Delaunay tetrahedralization of a sparse Structure-from-Motion point cloud.
As in~\cite{mai2026radiance}, we use a fixed outer sphere of vertices without parameters to close the triangulation.

Each vertex $k$ of the mesh stores a density (more precisely the logarithm of a density) and a feature vector $f_k$.
As in~\cite{mai2026radiance,govindarajan2025radfoam}, the density of each tetrahedron is constant within the tetrahedron, and this simplifies volume rendering.
We define the density of a tetrahedron as the geometric mean of the densities of its vertices.
Thus, a vertex with a density of 0 results in an "empty" tetrahedron with density 0.
Once the vertex positions freeze and the tetrahedra set no longer changes, we switch to per-tetrahedron densities following~\cite{mai2026radiance}.

\subsection{Rendering}
\label{sec:rasterization}
Volume rendering%
~\cite{volume_rendering_1988} integrates  features along a ray $r(t)$:
\begin{equation}
F = \int_0^\infty T(t)\,\sigma(r(t))\,f(r(t))\,dt \enspace ,
\label{eq:volrend}
\end{equation}
where, $r(t)$ is a 3D point on the ray, $\sigma(r(t))$ is the density at this point, $f(r(t))$ is a feature vector at this point, and $T(t) = \exp(-\int_0^t \sigma(s)\,ds)$ is the transmittance up to position $t$ along the ray.

A camera ray crosses a sequence of $N$ sorted tetrahedra. Each crossing is a segment $[t^i_\mathrm{in}, t^i_\mathrm{out}]$ of length $\ell_i$, $i = 1, \dots, N$.
By definition, the density within a segment is constant~\cite{govindarajan2025radfoam,mai2026radiance} and since we are using barycentric interpolation, the features are linear, so the integral over the segment has a closed form~\cite{mai2026radiance}.
Segment $i$ has opacity $\alpha_i$, transmittance $T_i$ in front of it and blending weight $w_i$:
\begin{equation}
\alpha_i = 1 - e^{-\sigma_i \ell_i}, \quad T_i = \prod_{j<i} (1 - \alpha_j), \quad w_i = T_i \alpha_i .
\label{eq:alpha}
\end{equation}
On segment $i$, the features vary linearly from $f_{\mathrm{in},i}$ at the entry point to $f_{\mathrm{out},i}$ at the exit point, both interpolated barycentrically from the vertices of the tetrahedron.
The integral of the features over the segment is~\cite{mai2026radiance}:
\begin{align}
F_i &= \int_{t^i_\mathrm{in}}^{t^i_\mathrm{out}} e^{-\sigma_i (t - t^i_\mathrm{in})}\, \sigma_i\, f(r(t))\, dt \nonumber \\
&= (1 - \varphi_i)\, f_{\mathrm{in},i} + (\varphi_i - e^{-\sigma_i \ell_i})\, f_{\mathrm{out},i} \enspace ,
\label{eq:segment}
\end{align}
with $\varphi_i = \alpha_i / (\sigma_i \ell_i)$.
Inside segment $i$, the transmittance is $T(t) = T_i\, e^{-\sigma_i (t - t^i_\mathrm{in})}$, so the integral of \cref{eq:volrend} is the sum of the segment integrals, each scaled by the transmittance in front of its segment,
\begin{equation}
F = \sum_{i=1}^{N} T_i\, F_i \enspace .
\label{eq:sum}
\end{equation}
See the supp. material for details.

As in~\cite{mai2026radiance}, we render the volumetric mesh via rasterization.
We sort the tetrahedra by the power distance of the camera to each circumsphere, which gives the exact visibility order of a Delaunay mesh~\cite{karasick1997visualization,mai2026radiance}. After each position update, the points are retriangulated, so the mesh maintains its Delaunay properties (up to numerical precision).
As in~\cite{kerbl3Dgaussians,mai2026radiance,govindarajan2026powerfoam},
the screen is split into tiles.
Each tetrahedron is binned to the tiles it covers.
Tetrahedra are rendered in front-to-back order, providing the intersections of a pixel's ray with the tetrahedra.
Each intersection defines one segment with its entry and exit features.
The rendering of a tile stops when none of its pixels has transmittance left.
Tetrahedra and segments with opacity below 0.001 are skipped, which increases efficiency (their gradients do not have to be computed and updated) and drops only very faint segments, \eg in the empty space.
The supp. mat. provides details on the
exact tile test, the skip test and the gradients.

To obtain the final color for a pixel, we use a small MLP to decode the ray's accumulated feature $F$ from~\cref{eq:sum} and a direction encoding $\gamma(r)$:
\begin{equation}
c = \mathrm{MLP}\big(F, \gamma(r)\big) \enspace .
\label{eq:color}
\end{equation}
We represent $\gamma$ using Spherical Harmonics (SH) as in instant-NGP~\cite{mueller2022instant} for the Mip-NeRF~360 outdoor scenes and using Fourier features as in Tetra-NeRF~\cite{kulhanek2023tetranerf} on all other scenes (see our two configurations in \cref{sec:exp}).
There is no separate background color.
A ray that does not intersect any tetrahedra is simply a zero feature.

\subsection{Optimization}
\label{sec:optimization}

We train for 30k iterations.
Until 18k iterations, the vertex positions are optimized.
Following~\cite{mai2026radiance},
the vertex positions are updated and the mesh is retriangulated every 10 iterations.
Between iterations 2k and 7.5k, vertices are added until a fixed vertex budget is reached.
After 18k iterations, the mesh is fixed, and the density of each tetrahedron is no longer tied to its vertices.
Each tetrahedron keeps the geometric mean of its vertex densities as its own density and continues to train it independently, which gives it more freedom now that the tetrahedra no longer change.
The last 12k iterations train only the features, the densities, and the MLP.

\noindent \textbf{Initialization.}
We initialize the mesh
by Delaunay triangulation of SfM points and adding the outer sphere~\cite{mai2026radiance}.
Densities are initialized as $e^{-4} \approx 0.018$ (almost transparent), and features are sampled from
$\mathcal{N}(0, 0.1^2)$.

\noindent \textbf{Losses.} We use the photometric loss used in~\cite{mai2026radiance}
\begin{equation}
\mathcal{L}_\mathrm{rgb} = (1 - \lambda_\mathrm{SSIM})\, \mathcal{L}_1 + \lambda_\mathrm{SSIM}\, (1 - \mathrm{SSIM})
\label{eq:loss}
\end{equation}
with $\lambda_\mathrm{SSIM}$ set to either 0.1 or 0.2.

\noindent \textbf{Faint-segment regularizer.} Volume rendering can explain a pixel by many faint segments spread along its ray, instead of a few opaque ones.
Skipping every segment below the skip threshold of 0.001
saves compositing time in the forward and backward pass.
To speed up training, from iteration 10k on, we push faint segments towards transparent so that more of them fall below the skip threshold.
Ideally, we would want each segment to either be empty or clearly be part of a surface.
We add a regularization term on the segments whose opacity is below a threshold $\theta_\mathrm{f}$.
For the ray of each pixel, with segment opacities $\alpha_1, \dots, \alpha_N$ (\cf \cref{eq:alpha}), the corresponding regularization term is
\begin{equation}
\mathcal{L}_\mathrm{faint} = \lambda_\mathrm{f} \sum_{i:\, \alpha_i < \theta_\mathrm{f}} \frac{4\, \alpha_i (\theta_\mathrm{f} - \alpha_i)}{\theta_\mathrm{f}^2} \enspace .
\label{eq:faint}
\end{equation}
Each summand
is a parabola that is 0 at $\alpha_i = 0$ and $\alpha_i = \theta_\mathrm{f}$ and peaks at $\theta_\mathrm{f}/2$.
The factor 4 scales this peak to 1, so the strength is set by the regularization weight $\lambda_\mathrm{f}$ alone.
The term pushes a segment below $\theta_\mathrm{f}/2$ towards transparent, until it falls below the skip threshold, and a segment above $\theta_\mathrm{f}/2$ up to $\theta_\mathrm{f}$.
Segments above $\theta_\mathrm{f}$ are not affected.
A stronger regularizer shortens training, but reduces rendering quality as more segments are skipped.
Thus, $\lambda_\mathrm{f}$ trades speed for quality (\cf supp. mat. for details).
We use $\theta_\mathrm{f} = 0.2$ and $\lambda_\mathrm{f} = 3 \cdot 10^{-4}$.
We add the mean of the regularization terms over all pixels in a training view to the loss.

\noindent \textbf{Positions.} Following~\cite{mai2026radiance}, the position gradients of the vertices accumulate over 10 iterations, and the positions take one step every 10 iterations.
Each step is followed by a retriangulation.
Retriangulating less and less frequently, as proposed in~\cite{govindarajan2025radfoam}, leads to about 20\% faster training times, at the costs of about 0.1\,dB.
We thus do not use this strategy.

\noindent \textbf{Densification.}
Similar to~\cite{kerbl3Dgaussians,govindarajan2025radfoam,govindarajan2026powerfoam,mai2026radiance}, we densify the volumetric mesh in regions that lead to inaccurate renderings by inserting vertices into tetrahedra.
Every 500 iterations between iterations 2k and 7.5k, we render 40 random training views and compute the per-pixel error $1 - \mathrm{SSIM}$.
Each tetrahedron collects the error of the pixels it contributes to, weighted by its blending weights.
A tetrahedron is a candidate for splitting when, in at least one view, it covers at least 2 pixels and its weighted average error is at least the average error over all pixels of the 40 views.
We then sample candidates with probability proportional to their error~\cite{mallick2024taming,govindarajan2026powerfoam} and add a vertex at the centroid of each selected tetrahedron.
Instead of selecting all tetrahedra with errors above some threshold (capped by another threshold for the maximum number of newly added vertices), each step adds as many vertices as needed to reach a target vertex count~\cite{mallick2024taming}.
At step $k$, the target is $n_0 (B / n_0)^{p_k}$, where $n_0$ is the initial number of vertices and $B$ the vertex budget.
The exponent $p_k$ grows from 0.30 at the first step to 1 at step 7.5k.
The steps thus add many vertices at first and fewer later.

New vertices should not be added into the free space in front of a surface.
We thus ignore the error of segments in front of the rendered surface, closer than 0.95 times the pixel's expected depth, normalized by its accumulated opacity.
We do this only for pixels whose accumulated opacity is above 0.5, where the expected depth can be considered as reliable.
A new vertex takes the mean features and the mean log density of the vertices of its tetrahedron, \ie the geometric mean of their densities.

\providecommand{\tbd}{\textcolor{red}{TBD}}
\providecommand{\old}[1]{\textcolor{gray}{#1}}

\section{Experiments}
\label{sec:exp}

\noindent \textbf{Setup.}
We evaluate SimpleTet on all 9 Mip-NeRF~360~\cite{barron2022mipnerf360} scenes and all 21 Tanks and Temples~\cite{knapitsch2017tanks} scenes, downsampled following common practice~\cite{kerbl3Dgaussians,govindarajan2025radfoam,mai2026radiance, kulhanek2024nerfbaselines}.
As in Radiance Meshes, we report Mip-NeRF~360 outdoor and indoor separately.
We use NerfBaselines~\cite{kulhanek2024nerfbaselines} to ensure a fair comparison between the methods.
All methods train on one A100-40GB, and training times exclude evaluation.

\begin{table}[t]
\centering
\caption{We use two configurations for our experiments. Fourier uses four frequencies from 1 to 2 cycles for direction encoding.
Settings not included in this table, \eg, the color MLP with 2 hidden layers of 256 dim., are shared between both configurations. See the supp. mat. for a complete list of all hyperparameters.}
\label{tab:settings}
\scriptsize
\setlength{\tabcolsep}{2.5pt}
\begin{tabular}{@{}l cc@{}}
\toprule
 & Outdoor config. & Default config. \\
 & (M360 outdoor) & (all other scenes) \\
\midrule
Vertex budget & 5M & 3M \\
Feature dim & 16 & 32 \\
Direction enc. & SH, degree 3 & Fourier \\
Position lr & $2 \cdot 10^{-4} \to 2 \cdot 10^{-6}$ & $10^{-4} \to 10^{-6}$ \\
Final feature lr & $3 \cdot 10^{-4}$ & $3 \cdot 10^{-3}$ \\
$\lambda_\mathrm{SSIM}$ & 0.1 & 0.2 \\
\bottomrule
\end{tabular}
\end{table}

\noindent \textbf{Implementation details.}
We use
gDel3D~\cite{cao2014gdel3d} for efficient GPU-based Delaunay triangulation.
We use it without its CPU-based star splaying, which only repairs small errors in the triangulation (\cf supp. mat.).
The supp. mat. details all our hyperparameter choices.
Like all baselines, our method renders one ray per pixel.
Similar to~\cite{Held2025MeshSplatting}, we also experiment with test-time $2\times2$ supersampling (SS).

\noindent \textbf{Baselines.} We compare with Radiance Meshes~\cite{mai2026radiance}, PowerFoam~\cite{govindarajan2026powerfoam}, Radiant Foam~\cite{govindarajan2025radfoam} and VoroTracing~\cite{taveira2026vorotracing}.
The methods are run inside NerfBaselines~\cite{kulhanek2024nerfbaselines} to ensure identical and fair evaluation conditions.
We report both the results obtained via NerfBaselines and the results reported by the authors (in gray).
The supp. mat. provides a detailed comparison between the two sets of results.

We use the official configurations for each method: Radiant Foam and VoroTracing are trained for 20k iterations, while all other methods are trained for 30k iterations.
Radiance Meshes uses one configuration for all scenes.
PowerFoam, Radiant Foam, and VoroTracing use the  indoor and outdoor configurations they provide for Mip-NeRF~360.
On Tanks and Temples, we run each of them with its default and its matching Mip-NeRF~360 configuration per scene type and report the better one (see the supp.\ mat. for details).

\noindent \textbf{Outdoor and default configurations.}
Similar to~\cite{govindarajan2026powerfoam,govindarajan2025radfoam,taveira2026vorotracing}, we use two configurations, an outdoor configuration for the unbounded Mip-NeRF~360 outdoor scenes and a default configuration for all other scenes.
\cref{tab:settings} shows the differences between both configurations.
As is common in the literature, the outdoor configuration has a larger vertex budget.
Compared to indoor scenes, outdoor scenes tend to span a larger area and often have few SfM points in regions distant from the training cameras.
Thus, vertices in such parts of the scene need to be able to move larger distances during optimization.
Hence, we double the position learning rate for outdoor scenes.
The default configuration encodes the view direction with the Fourier features of Tetra-NeRF~\cite{kulhanek2023tetranerf}, with frequencies from 1 to 2 cycles.
(instead of the 1 to 16 cycles used in~\cite{kulhanek2023tetranerf}).
The outdoor configuration uses Spherical Harmonics (SH) of degree 3 instead, similar to~\cite{mueller2022instant,tancik2023nerfstudio}.
We use $\lambda_\mathrm{SSIM} = 0.1$ for the outdoor and $\lambda_\mathrm{SSIM} = 0.2$ for the default configuration.

\subsection{Comparison with state-of-the-art methods}
\label{sec:comparison}

\begin{table*}[t!h]
\centering
\caption{\textbf{Comparison with state-of-the-art methods using volumetric mesh-based scene representations}. Per group of scenes, we report the means over their scenes. We also report the means over all scenes (``All 30''), and the means obtained by averaging the the Mip-NeRF~360 means (9 scenes) and the Tanks and Temples  (T\&T) means (``Mean of both datasets''). Training times are averaged over all 30 scenes.
We highlight \colorbox{rkfirst}{first}, \colorbox{rksecond}{second} and \colorbox{rkthird}{third} best results among the baselines and our SimpleTet approach.
Besides the results obtained by using exactly the same evaluation protocol for all methods, we show the results reported for the baselines in their corresponding publications in gray (not taken into account for the ranking; we only show numbers if the method was evaluated on all scenes in a group).
We further report variants of SimpleTet using test-time supersampling (+SS), and using the same numbers of vertices as~\cite{mai2026radiance}  (RM \#vert.).
\rfmark{}~We use gDel3D for triangulation for \cite{govindarajan2025radfoam} on the Train scene from T\&T, as their original Delaunay implementation crashes for this scene.}
\label{tab:nvs}
\footnotesize
\setlength{\tabcolsep}{1.85pt}
\begin{tabular}{@{}l ccc ccc ccc ccc ccc c@{}}
\toprule
 & \multicolumn{3}{c}{M360 outdoor (5)} & \multicolumn{3}{c}{M360 indoor (4)} & \multicolumn{3}{c}{Tanks and Temples (21)} & \multicolumn{3}{c}{All 30 scenes} & \multicolumn{3}{c}{Mean of both datasets} & Train \\
\cmidrule(lr){2-4}\cmidrule(lr){5-7}\cmidrule(lr){8-10}\cmidrule(lr){11-13}\cmidrule(lr){14-16}
Method & PSNR & SSIM & LPIPS & PSNR & SSIM & LPIPS & PSNR & SSIM & LPIPS & PSNR & SSIM & LPIPS & PSNR & SSIM & LPIPS & min \\
\midrule
Radiance Meshes~\cite{mai2026radiance}          & \fst{24.59} & \fst{0.726} & \snd{0.203} & 30.15 & \snd{0.918} & 0.130 & \snd{23.16} & \snd{0.822} & \snd{0.166} & \snd{24.33} & \snd{0.819} & \snd{0.167} & \trd{25.11} & \snd{0.816} & \snd{0.168} & 320 \\
\pap{\quad as reported} & \pap{24.38} & \pap{0.721} & \pap{--} & \pap{30.61} & \pap{0.920} & \pap{--} & \pap{--} & \pap{--} & \pap{--} & \pap{--} & \pap{--} & \pap{--} & \pap{--} & \pap{--} & \pap{--} & \pap{--} \\
Radiance Meshes w/o culling & 24.53 & 0.722 & 0.207 & 30.58 & 0.919 & 0.127 & 23.32 & 0.823 & 0.163 & 24.49 & 0.819 & 0.166 & 25.27 & 0.816 & 0.167 & 386 \\
PowerFoam~\cite{govindarajan2026powerfoam} & 23.96 & 0.669 & 0.222 & \trd{31.25} & 0.912 & \trd{0.122} & 22.68 & 0.756 & 0.241 & 24.03 & 0.763 & 0.222 & 24.94 & 0.767 & 0.209 & \fst{40} \\
\pap{\quad as reported} & \pap{--} & \pap{--} & \pap{--} & \pap{31.45} & \pap{0.918} & \pap{--} & \pap{--} & \pap{--} & \pap{--} & \pap{--} & \pap{--} & \pap{--} & \pap{--} & \pap{--} & \pap{--} & \pap{--} \\
Radiant Foam~\cite{govindarajan2025radfoam} & 23.93 & 0.669 & 0.256 & 30.67 & 0.901 & 0.127 & 22.17\rfmark{} & 0.788\rfmark{} & 0.184\rfmark{} & 23.59\rfmark{} & 0.783\rfmark{} & 0.188\rfmark{} & 24.55\rfmark{} & 0.780\rfmark{} & 0.191\rfmark{} & 149\rfmark{} \\
\pap{\quad as reported} & \pap{--} & \pap{--} & \pap{--} & \pap{30.75} & \pap{0.908} & \pap{--} & \pap{--} & \pap{--} & \pap{--} & \pap{--} & \pap{--} & \pap{--} & \pap{--} & \pap{--} & \pap{--} & \pap{--} \\
VoroTracing~\cite{taveira2026vorotracing} & \trd{24.28} & \trd{0.695} & \trd{0.209} & \snd{31.35} & \trd{0.915} & \snd{0.112} & \trd{22.96} & \trd{0.789} & \trd{0.183} & \trd{24.30} & \trd{0.790} & \trd{0.178} & \snd{25.19} & \trd{0.791} & \trd{0.175} & \trd{75} \\
\pap{\quad as reported} & \pap{--} & \pap{--} & \pap{--} & \pap{31.42} & \pap{0.916} & \pap{--} & \pap{--} & \pap{--} & \pap{--} & \pap{--} & \pap{--} & \pap{--} & \pap{--} & \pap{--} & \pap{--} & \pap{--} \\
\midrule
Ours                     & \snd{24.42} & \snd{0.718} & \fst{0.188} & \fst{32.24} & \fst{0.930} & \fst{0.106} & \fst{24.48} & \fst{0.838} & \fst{0.125} & \fst{25.51} & \fst{0.830} & \fst{0.133} & \fst{26.19} & \fst{0.825} & \fst{0.138} & \snd{43} \\
Ours + SS                & 24.62 & 0.724 & 0.200 & 32.31 & 0.932 & 0.110 & 24.60 & 0.842 & 0.135 & 25.63 & 0.834 & 0.142 & 26.32 & 0.829 & 0.147 & 43 \\
Ours, RM \#vert.         & 24.08 & 0.693 & 0.224 & 32.05 & 0.927 & 0.114 & 24.49 & 0.836 & 0.137 & 25.43 & 0.824 & 0.149 & 26.06 & 0.816 & 0.156 & 26 \\
Ours, RM \#vert. + SS    & 24.24 & 0.698 & 0.236 & 32.12 & 0.928 & 0.118 & 24.59 & 0.838 & 0.148 & 25.54 & 0.827 & 0.159 & 26.17 & 0.819 & 0.166 & 26 \\
\bottomrule
\end{tabular}
\end{table*}

\Cref{tab:nvs} compares our SimpleTet approach to current state-of-the-art methods on the Mip-NeRF~360 and Tanks and Temples (T\&T) datasets.
As can be seen, our approach performs similarly or better compared to all baselines in terms of novel view synthesis quality.
Our approach is significantly faster to train than Radiance Meshes, which is overall the best-performing baseline in terms of rendering quality.
At the same time, SimpleTet clearly outperforms PowerFoam, the baseline with the lowest training time, under all image quality metrics while being only slightly less efficient to train.
Test-time supersampling (+SS) improves rendering quality in terms of PSNR and SSIM, at 4x slower rendering speeds.
At the same time, supersampling smooths out
the rendered image, resulting in a slight decrease in terms of LPIPS.
Unlike our method, Radiance Meshes uses culling of nearly transparent tetrahedra during training.
Without culling, novel view synthesis quality improves at significantly longer training times.
Still, our method outperforms Radiance Meshes without culling on most scenes.

\begin{table}[t]
\centering
\caption{\textbf{Model size, training time, and training memory}, averaged over all 30 scenes. Params: stored float values of the trained model, without connectivity. GPU mem.: peak GPU memory of the training process as recorded by NerfBaselines. $^a$Trained with a hash grid (427M parameters), baked to per-tetrahedron values after 18k training iterations, 5.5 tetrahedra per vertex. $^b$3 position and 16 or 32 feature values per vertex (depending on the type of scene),
plus one density per tetrahedron, 6.5 tetrahedra per vertex. \rfmark{}Using our triangulation for the T\&T Train scene.}%
\label{tab:size}
\scriptsize
\setlength{\tabcolsep}{2pt}
\begin{tabular}{@{}l l rrr rr@{}}
\toprule
Method & Primitive & \#prim. & values & params & train & GPU mem. \\
 & & (M) & / prim. & (M) & (min) & (GB) \\
\midrule
Radiance Meshes & Delaunay vertex$^a$ & 1.27 & 52 / tet & 365 & 320 & 14.1 \\
PowerFoam       & power-diagram site & 0.62 & 417 & 257 & 40 & 19.7 \\
Radiant Foam    & Voronoi site       & 2.35 & 52  & 122 & 149\rfmark{} & 30.1 \\
VoroTracing     & Voronoi site       & 2.00 & 388 & 776 & 75 & 20.9 \\
\midrule
Ours            & Delaunay vertex    & 3.33 & 19 / 35$^b$ & 125 & 43 & 14.9 \\
Ours, RM \#vert. & Delaunay vertex & 1.27 & 19 / 35$^b$ & 48 & 26 & 9.3 \\
\bottomrule
\end{tabular}
\end{table}

Radiance Meshes uses fewer vertices per scene compared to our method (0.45--2.0M).
\Cref{tab:nvs} thus also reports results for SimpleTet using the same number of vertices per scene as Radiance Meshes.
As can be seen, our approach still outperforms Radiance Meshes on all but the Mip-NeRF~360 outdoor scenes.
At the same time, using fewer vertices significantly reduces training times.

Vertex count alone is a poor measure of model size and complexity, since different methods store different numbers of attributes per primitive.
\Cref{tab:size} thus compares training time, peak memory consumption during training, and model size.
We train faster than all baselines except PowerFoam.
At the vertex count of Radiance Meshes, we train faster than PowerFoam while still being better in all three metrics (\cf \Cref{tab:nvs}).
Training requires less memory than all baselines except Radiance Meshes. When our method matches the vertex count of Radiance Meshes, we use the least memory.

The number of parameters reported in \Cref{tab:size} does not account for the 427M parameters used by the instant-NGP model used by Radiance Meshes as part of its training process.
Still, the number of parameters of our approach is similar or smaller compared to all baselines.

\begin{figure*}[t]
\centering
\includegraphics[width=0.9\linewidth]{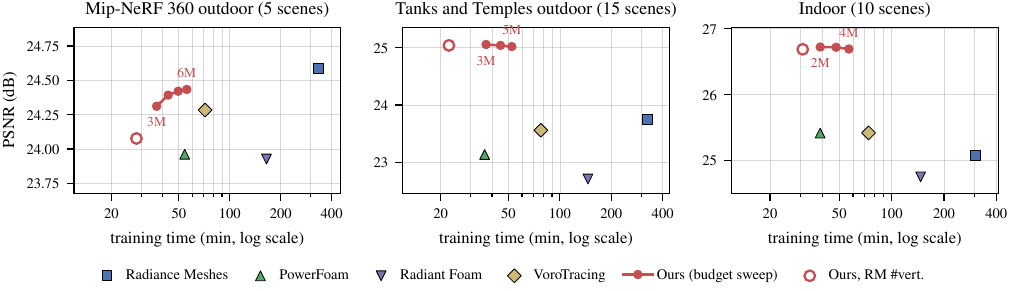}
\caption{\textbf{PSNR \vs training time (log scale).} We report mean PSNR values per dataset.
We vary the vertex budget for our approach to obtain a curve. The open red circle denotes the performance of our approach using the vertex budget of Radiance Meshes (RM)~\cite{mai2026radiance}. As can be seen, our approach offers the best combination of rendering quality and training time.}
\label{fig:budget}
\end{figure*}

\Cref{fig:budget} shows the trade-off between PSNR and training time when varying the number of vertices used by SimpleTet.
As can be seen, the number of vertices has little impact on PSNR for the Tanks and Temples and Indoor scenes.
Using more vertices help the most on the unbounded Mip-NeRF~360 outdoor scenes.
We observe a similar behavior for SSIM and LPIPS (\cf supp. mat.).
Overall, our approach offers the best trade-off between novel view synthesis quality and training times among all methods.

\begin{figure*}[t]
\centering
\includegraphics[width=0.9\linewidth]{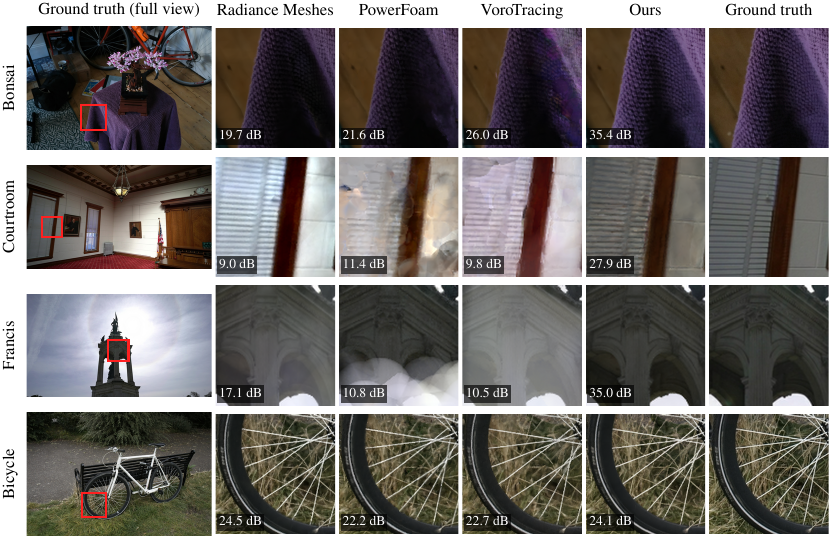}
\caption{\textbf{Crops from test views,} with the local PSNR of each crop.
Left: the full ground-truth image with the crop box. The crops are chosen automatically: on Bonsai, Courtroom, and Francis among the windows where SimpleTet improves most over the best baseline, on Bicycle the window with the median difference to the best baseline. On Courtroom and Francis much of the baselines' error is a shift in brightness or color.}
\label{fig:qual}
\end{figure*}

\Cref{fig:qual} shows qualitative results for the baselines and our approach, illustrating the visual quality of SimpleTet.

\subsection{Ablations}
\label{sec:ablations}

In a second series of experiments, we analyze the impact of various design choices on the performance of our method.

\begin{table}[t]
\centering
\caption{\textbf{Validating vertex movement and densification.} ``fixed'' keeps the mesh of Tetra-NeRF, ``moving'' adds our vertex movement, and ``1M vertices'' uses our densification, starting from the SfM points, instead of the 1M points provided by Tetra-NeRF~\cite{kulhanek2023tetranerf}.}
\label{tab:tetranerf}
\scriptsize
\setlength{\tabcolsep}{2pt}
\resizebox{\columnwidth}{!}{%
\begin{tabular}{@{}l cc cc ccc c@{}}
\toprule
 & & & Outdoor (5) & Indoor (4) & \multicolumn{3}{c}{Mip-NeRF~360 (9)} & Train \\
\cmidrule(lr){4-4}\cmidrule(lr){5-5}\cmidrule(lr){6-8}
Method & \#vert. & Iter. & PSNR & PSNR & PSNR & SSIM & LPIPS & min \\
\midrule
Tetra-NeRF & 1M & 300k & 23.36 & 30.21 & 26.40 & 0.727 & 0.269 & 1199 \\
\midrule
Ours, Tetra-NeRF points, fixed & 1M & 30k & 23.65 & 31.01 & 26.92 & 0.779 & 0.216 & 31 \\
Ours, Tetra-NeRF points, moving & 1M & 30k & 24.16 & 31.56 & 27.45 & 0.798 & 0.199 & 32 \\
Ours, 1M vertices & 1M & 30k & 23.91 & 32.02 & 27.52 & 0.788 & 0.190 & 31 \\
\midrule
Ours & 4.1M & 30k & 24.42 & 32.24 & 27.90 & 0.812 & 0.152 & 56 \\
\midrule
Zip-NeRF~\cite{barron2023zipnerf} & -- & 200k & 25.57 & 32.28 & 28.55 & 0.829 & 0.138 & 330 \\
3DGS-MCMC~\cite{kheradmand2024mcmc} & -- & 30k & 25.18 & 31.49 & 27.98 & 0.835 & 0.148 & 41 \\
\bottomrule
\end{tabular}}
\end{table}

\noindent \textbf{Validating densification and vertex position optimization.}
\cref{tab:tetranerf} shows the importance of optimizing vertex positions and densifying the underlying volumetric mesh.
As a baseline, we use dense point clouds, each containing 1M vertices, reconstructed using COLMAP~\cite{schoenberger2016mvs, Schonberger_2016_CVPR} and provided as input for Tetra-NeRF~\cite{kulhanek2023tetranerf}.
\cref{tab:tetranerf} shows that optimizing the vertex positions during training improves reconstruction quality.
Our densification approach, starting from sparse SfM points, achieves a similar performance as the dense point clouds.
Combining both vertex optimization and densification clearly leads to the best results.
For reference, we include Tetra-NeRF results using the same dense point clouds since Tetra-NeRF keeps the vertices fixed and does not perform densification.
\cref{tab:tetranerf} further shows that SimpleTet approaches the rendering quality of state-of-the-art NeRF and 3DGS approaches (see also supp. mat.).

\begin{table}[t]
\centering
\caption{\textbf{Ablating design choices.} Ablations, one ray per pixel.
We average results over the 9 Mip-NeRF~360 and the 21 Tanks and Temples scenes. The rows below SimpleTet show the differences in performance when changing different parts of the method. ``minimum'' runs in PyTorch~\cite{Ansel_PyTorch_2_Faster_2024} instead of our CUDA kernel, which slows down training. The default MLP width is 256 (2 hidden layers). We also show  results for spherical harmonics (SH) of degree 3 with coefficients per vertex instead of features and MLP. }
\label{tab:abl}
\scriptsize
\setlength{\tabcolsep}{1.6pt}
\resizebox{\columnwidth}{!}{%
\begin{tabular}{@{}l ccc ccc c@{}}
\toprule
 & \multicolumn{3}{c}{Mip-NeRF~360 (9)} & \multicolumn{3}{c}{T\&T (21)} & Train \\
\cmidrule(lr){2-4}\cmidrule(lr){5-7}
Method & PSNR & SSIM & LPIPS & PSNR & SSIM & LPIPS & min \\
\midrule
SimpleTet & 27.90 & 0.812 & 0.152 & 24.48 & 0.838 & 0.125 & 43 \\
\midrule
+ SS & +0.14 & +0.004 & +0.008 & +0.12 & +0.004 & +0.009 & \old{+0} \\
\midrule
No faint-segment regularizer & \old{+0.03} & \old{+0.001} & \old{+0.001} & +0.06 & +0.002 & \old{0.000} & +4 \\
\midrule
Features: hash grid at vertices & $-$0.79 & $-$0.009 & +0.011 & $-$0.85 & $-$0.014 & +0.025 & +105 \\
No per-tetrahedron densities & $-$0.06 & $-$0.003 & +0.002 & $-$0.11 & $-$0.002 & +0.004 & \old{$-$2} \\
Density: arithmetic mean & $-$0.41 & $-$0.011 & +0.009 & $-$0.46 & $-$0.011 & +0.015 & \old{+1} \\
Density: minimum & $-$1.45 & $-$0.037 & +0.032 & $-$1.88 & $-$0.065 & +0.067 & +14 \\
\midrule
Color: MLP width 128 & $-$0.16 & $-$0.002 & +0.002 & $-$0.11 & $-$0.002 & +0.003 & \old{$-$2} \\
Color: MLP width 64 & $-$0.27 & $-$0.004 & +0.004 & $-$0.25 & $-$0.006 & +0.008 & \old{$-$3} \\
Color: MLP width 32 & $-$0.57 & $-$0.009 & +0.009 & $-$0.51 & $-$0.010 & +0.015 & \old{$-$3} \\
Color: SH instead of MLP & $-$0.73 & $-$0.004 & +0.009 & $-$1.08 & $-$0.013 & +0.020 & +9 \\
Color: SH, colors accumulated & $-$1.03 & $-$0.017 & +0.024 & $-$1.46 & $-$0.027 & +0.044 & +39 \\
Color: other direction encoding & $-$0.28 & $-$0.006 & +0.005 & \old{$-$0.01} & \old{$-$0.001} & +0.003 & \old{+1} \\
Color: SH directions on all scenes & $-$0.15 & \old{$-$0.001} & \old{+0.001} & \old{$-$0.01} & \old{$-$0.001} & +0.003 & \old{+2} \\
Color: Fourier directions on all scenes & $-$0.12 & $-$0.006 & +0.005 & \old{0.00} & \old{0.000} & \old{0.000} & \old{0} \\
\bottomrule
\end{tabular}}
\end{table}

\noindent \textbf{Design choices.}
\Cref{tab:abl} ablates key design choices for our method on all 30 scenes, in the order of \cref{sec:method}.
Most differences are small. For reference, rerunning our method changes a single scene by at most 0.03\,dB PSNR, 0.001 SSIM and LPIPS, and 3\,min of training time.

\noindent \textbf{Hash-grid instead of per-vertex features.}
\cite{mai2026radiance} argue that using an instant-NGP hash grid~\cite{mueller2022instant} to store densities and features outperforms per-vertex features and densities.
We thus compare SimpleTet with a variant that uses an instant-NGP hash grid~\cite{mueller2022instant}, with the grid, heads and learning rates of Radiance Meshes~\cite{mai2026radiance}.
\Cref{tab:abl} observes that in our pipeline, the grid performs worse for every one of the 30 scenes.
We notice that a retriangulation still changes which vertices are interpolated at a point, so the grid does not remove the discontinuities of the edge flips, although the values at neighboring vertices may be smoother compared to per-vertex features.
We also measure the edge flips directly on bicycle, garden and room (\cf supp.\ mat.).
A retriangulation costs only 0.01--0.09\,dB PSNR.
Thus, edge flips do not seem to
significantly disturb the training process.

\noindent \textbf{Per-tetrahedron density representation.}
As can be seen from \cref{tab:abl}, the geometric mean of the vertex densities is better than their arithmetic mean on every scene, by 0.44\,dB on average, and far better than their minimum.
With the arithmetic mean, a dense vertex increases the density of all its adjacent tetrahedra, which makes it harder to model free space next to the surface.
With the minimum, the gradient of a tetrahedron's density only affects the vertex with minimum density.
In contrast, the geometric mean enables one ``empty'' vertex (a vertex with density 0) to drop the density of a tetrahedron to 0, and spreads the gradient over all four vertices. The supp.\ mat.\ provides additional details on how different density representations behave.

The large impact of the density representation on novel view synthesis quality might explain why~\cite{mai2026radiance} observe worse results for per-vertex features and densities.\footnote{\cite{mai2026radiance} does not seem to provide details on how the per-vertex values influence the tetrahedra.}

The switch to per-tetrahedron densities after 18k iteration
gains 0.1\,dB over the 30 scenes at the same training time, on 23 of the 30 scenes, with better SSIM and LPIPS (\cf ``No per-tetrahedron densities'' in  \cref{tab:abl}).

\noindent \textbf{Color representation.}
\cref{tab:abl} clearly shows the advantages of using an MLP to decode color from accumulated features:
Using per-vertex Spherical Harmonics (SH) colors that are interpolated within tetrahedra and accumulated along the ray
clearly decreases performance (``Color: SH, colors accumulated'').
Accumulating SH coefficients as our features and evaluating them once per pixel increases performance, but still performs worse than using an MLP (``Color: SH instead of MLP'').

Hahlbohm \etal~\cite{hahlbohm2026compact} also replace the SH coefficients of 3DGS by an MLP, but decode a feature per Gaussian before blending, with 2 hidden layers of width 16.
We decode once per pixel mainly for efficiency.
The MLP runs outside the rasterization kernel and its cost does not grow with the number of segments, so using a wider MLP improves performance at limited training time increases.

We use different direction encodings for our two configurations (\cf \cref{tab:settings}).
While using SH to encode directions is important for the Mip-NeRF~360 outdoor scenes, there is little difference in using SH instead of the Fourier representation for all other scenes (\cf the last three lines in \cref{tab:abl}).

\subsection{Limitations}
\label{sec:limitations}
Unbounded backgrounds and foliage are hard to model for all methods.
On Mip-NeRF~360, every method has its lowest PSNR on the Flowers and Treehill scenes.
Radiance Meshes is better than SimpleTet in terms of PSNR on Bicycle, Flowers and Treehill from Mip-NeRF~360, and on Truck from Tanks and Temples.
Away from the training views, all methods show floaters and a broken-up far background.
The supp.\ mat.\ visualizes these failure cases.

Enforcing that tetrahedra are either empty or occupied is desirable as empty tetrahedra can be removed, thus increasing rendering speed and decreasing the memory footprint.
Currently, SimpleTet does not enforce this behavior.

\section{Conclusion}
This paper proposed SimpleTet, a conceptually simple approach
using a tetrahedral scene representation.
Contrary to claims in the literature, we show that a simple representation that stores features and densities at the vertices of the tetrahedra can achieve state-of-the-art performance, both in terms of rendering quality and training efficiency.
Detailed ablation studies show that properly modeling per-tetrahedron densities and the color function has a significant impact on  the effectiveness and efficiency of our method.
We also show that handling discontinuities, which can arise when using tetrahedral meshes as a scene representation, is not always necessary in practice, despite prior work devising complex approaches to address this issue.

\PAR{Acknowledgements}
This work was supported by the Czech Science Foundation (GA\v{C}R) EXPRO (grant no. 23-07973X),
and by the Ministry of Education, Youth and Sports of the Czech Republic through the e-INFRA CZ (ID:90254).

{
    \small
    \bibliographystyle{ieeenat_fullname}
    \bibliography{main}
}

\clearpage
\setcounter{page}{1}
\maketitlesupplementary

\setcounter{section}{0}
\setcounter{table}{0}
\setcounter{figure}{0}
\renewcommand{\thesection}{S\arabic{section}}
\renewcommand{\thetable}{S\arabic{table}}
\renewcommand{\thefigure}{S\arabic{figure}}
\renewcommand{\theHsection}{S\arabic{section}}
\renewcommand{\theHtable}{S\arabic{table}}
\renewcommand{\theHfigure}{S\arabic{figure}}
\renewcommand{\floatpagefraction}{0.6}
\renewcommand{\dblfloatpagefraction}{0.6}
\raggedbottom

This supplementary material provides further details and results.
\Cref{sec:supp_method} details the rendering of \cref{sec:rasterization} and the components that we tested but do not use.
\Cref{sec:supp_setup} describes the datasets, the configurations of the baselines and all hyperparameters of the setup used in \cref{sec:exp}.
\Cref{sec:supp_results} extends the comparison of \cref{sec:comparison} with results per scene type and per scene, a comparison with NeRF~\cite{mildenhall2020nerf} and Gaussian splatting~\cite{kerbl3Dgaussians} methods and the vertex budget, and shows the failure cases of \cref{sec:limitations}.
\Cref{sec:supp_efficiency} breaks down the training time and details the rendering speed.
\Cref{sec:supp_ablations} extends the ablations of \cref{sec:ablations} and analyzes the faint-segment regularizer of \cref{sec:optimization}.

\section{Method details}
\label{sec:supp_method}

\paragraph{Densities.} Each vertex $k$ stores a log density $\rho_k$, except the vertices of the outer sphere, which have no density.
A tetrahedron $t$ has the geometric mean of the densities $e^{\rho_k}$ of its vertices $V_t$, where $V_t$ leaves out the vertices of the outer sphere,
\begin{equation}
\sigma_t = \Big(\prod_{k \in V_t} e^{\rho_k}\Big)^{1/|V_t|} = \exp\Big(\frac{1}{|V_t|} \sum_{k \in V_t} \rho_k\Big) ,
\label{eq:density}
\end{equation}
and a tetrahedron with only outer-sphere vertices has density 0.

\paragraph{Features.} At a point $x$ in tetrahedron $t$ with barycentric coordinates $\beta_k(x)$, the features are
\begin{equation}
f(x) = \sum_{k \in V_t} \beta_k(x)\, f_k \enspace .
\label{eq:feature}
\end{equation}
Segment $i$ enters tetrahedron $t$ at the point $x^i_\mathrm{in} = r(t^i_\mathrm{in})$ on one face and leaves it at $x^i_\mathrm{out} = r(t^i_\mathrm{out})$ on another face.
The entry and exit features of \cref{eq:segment} are the features at these two points,
\begin{equation}
f_{\mathrm{in},i} = f(x^i_\mathrm{in}) \enspace , \quad f_{\mathrm{out},i} = f(x^i_\mathrm{out}) \enspace .
\end{equation}
On a face, only the three vertices of that face have nonzero barycentric coordinates, so the exit features of a segment are the entry features of the next one.
In between, the features $f_i(s) = f(r(t^i_\mathrm{in} + s))$ change linearly with the distance $s \in [0, \ell_i]$ from the entry point, where $\ell_i = t^i_\mathrm{out} - t^i_\mathrm{in}$ is the length of the segment,
\begin{equation}
f_i(s) = \Big(1 - \frac{s}{\ell_i}\Big) f_{\mathrm{in},i} + \frac{s}{\ell_i}\, f_{\mathrm{out},i} \enspace .
\label{eq:supp_linear}
\end{equation}

\paragraph{Segment integral.} With $s = t - t^i_\mathrm{in}$, \cref{eq:segment} reads
\begin{equation}
F_i = \int_0^{\ell_i} \sigma_i e^{-\sigma_i s} f_i(s)\, ds \enspace .
\end{equation}
A thin slice of the segment at depth $s$ has the opacity $\sigma_i\, ds$ and is seen through the part of the segment in front of it, whose transmittance is $e^{-\sigma_i s}$.
So $\sigma_i e^{-\sigma_i s}\, ds$ is the weight of the slice, the weights of all slices sum to $\alpha_i$, and the front of the segment has a larger influence than its back.
Substituting the linear features of \cref{eq:supp_linear} splits $F_i$ into an entry and an exit term,
\begin{equation}
\begin{split}
F_i = {} & f_{\mathrm{in},i} \int_0^{\ell_i} \sigma_i e^{-\sigma_i s} \Big(1 - \frac{s}{\ell_i}\Big)\, ds \\
& + f_{\mathrm{out},i} \int_0^{\ell_i} \sigma_i e^{-\sigma_i s}\, \frac{s}{\ell_i}\, ds \enspace .
\end{split}
\end{equation}
Integration by parts gives the weight of the exit features,
\begin{equation}
\int_0^{\ell_i} \sigma_i e^{-\sigma_i s}\, \frac{s}{\ell_i}\, ds = \varphi_i - e^{-\sigma_i \ell_i} \enspace ,
\end{equation}
where
\begin{equation}
\varphi_i = \frac{1}{\ell_i} \int_0^{\ell_i} e^{-\sigma_i s}\, ds = \frac{\alpha_i}{\sigma_i \ell_i} \enspace .
\end{equation}
The two weights sum to $\alpha_i$, so the entry features get $\alpha_i - (\varphi_i - e^{-\sigma_i \ell_i}) = 1 - \varphi_i$, which gives \cref{eq:segment}.
Here $\varphi_i$ is the mean transmittance within the segment, the average of $e^{-\sigma_i s}$, which falls from 1 at the entry point to $e^{-\sigma_i \ell_i}$ at the exit point.
It is close to 1 for a nearly transparent segment and close to 0 for a nearly opaque one, and the lower it is, the more of $\alpha_i$ goes to the entry features.
For a thin or transparent segment, $\sigma_i \ell_i \to 0$, the transmittance $e^{-\sigma_i s} \approx 1 - \sigma_i s$ falls almost linearly, so its mean is its value at the middle of the segment, $\varphi_i \approx 1 - \sigma_i \ell_i / 2$, both weights are about $\sigma_i \ell_i / 2$, and $F_i$ is the mean of the entry and exit features times $\alpha_i \approx \sigma_i \ell_i$.
For an opaque segment, $\sigma_i \ell_i \to \infty$, we have $\varphi_i \to 0$, the weights tend to 1 and 0, and the ray sees only the entry features, $F_i \to f_{\mathrm{in},i}$.
For small $\sigma_i \ell_i$, the entry and exit weights $1 - \varphi_i$ and $\varphi_i - e^{-\sigma_i \ell_i}$ are differences of two numbers close to 1 and lose accuracy in single precision.
For $\sigma_i \ell_i < 0.25$, we therefore evaluate them with their Taylor series.

\paragraph{Blending weights.} By \cref{eq:segment,eq:sum}, segment $i$ adds
\begin{equation}
T_i F_i = T_i (1 - \varphi_i)\, f_{\mathrm{in},i} + T_i (\varphi_i - e^{-\sigma_i \ell_i})\, f_{\mathrm{out},i}
\end{equation}
to the pixel.
The blending weight $w_i = T_i \alpha_i$ of \cref{eq:alpha} is the share of the pixel that segment $i$ covers, its opacity $\alpha_i$ attenuated by the transmittance $T_i$ in front of it.
Its two weights sum to $w_i$, so $w_i$ is the total weight of segment $i$ in the pixel, split between its entry and exit features.
With constant features $f$, the contribution would be $w_i f$, as in alpha compositing.
With linear features, there is no single feature to multiply $w_i$ with, so \cref{eq:sum} is written with $T_i$, and $\alpha_i$ is inside $F_i$.
We use $w_i$ where one weight per segment is needed, as in the densification, where each tetrahedron collects the error of a pixel weighted by $w_i$.

\paragraph{Compositing.} The segments are composited front to back, starting with $T_1 = 1$ and $F = 0$, and $T_{i+1} = T_i (1 - \alpha_i)$.
Segments with $\alpha_i < 0.001$ are skipped, and, as in Radiance Meshes~\cite{mai2026radiance}, a tile stops once the transmittance of all its pixels is so low that the segments behind could change their color by at most about one 8-bit level.
The MLP then decodes $F$ once per pixel.

\paragraph{Outer sphere.} As in Radiance Meshes~\cite{mai2026radiance}, 1000 vertices on a Fibonacci sphere at the distance of the farthest SfM point. The vertices on the sphere have no parameters and are left out of $V_t$ in \cref{eq:density}. There is no Mip-NeRF~360~\cite{barron2022mipnerf360} style contraction of space, since we do not sample points along the ray or query a field, but intersect the ray with tetrahedra.

\paragraph{Direction encoding.} Our default configuration encodes each component of the view direction $d$ as
\begin{equation}
\gamma(d) = \big(d,\ \sin(2\pi f_j d),\ \cos(2\pi f_j d)\big)_{j=0}^{3} \enspace ,
\label{eq:direnc}
\end{equation}
with the frequencies $f_j = 2^{j/3}$, \ie four frequencies spaced evenly in log scale from 1 to 2 cycles.
Tetra-NeRF~\cite{kulhanek2023tetranerf} uses the same encoding with four frequencies from 1 to 16 cycles.
Our outdoor configuration  uses the 16 real spherical harmonics up to degree 3, evaluated at $d$, as instant-NGP~\cite{mueller2022instant}.
Ref-NeRF~\cite{verbin2022refnerf} encodes the reflection direction with spherical harmonics, damped by the roughness.
In preliminary experiments, the reflected direction did not help us, so we kept the view direction.

\paragraph{Tile binning.} On the screen, a tetrahedron covers a convex polygon, the convex hull of its four projected vertices.
As in the exact tile test of the 3DGS CUDA blog~\cite{kopanas2026cudablog}, we assign the tetrahedron only to the tiles that this polygon overlaps, not to all tiles of its bounding box.
We find these tiles row by row.
Because the polygon is convex, within one row of tiles it covers a single contiguous run of tiles, from its leftmost to its rightmost point in that row.
Both points are easy to find, since each of them is either a projected vertex that lies in the row, or a point where the line between two projected vertices crosses the top or bottom border of the row.
If a vertex lies behind the camera, we use the bounding box.
Each pixel then intersects its ray with the four face planes of each tetrahedron in its tile's list, which gives the entry and exit points, and a tile stops when none of its pixels has transmittance left.

\paragraph{Faint tetrahedra and segments.} A tetrahedron is not drawn when its largest possible opacity is below $\alpha_\mathrm{min} = 0.001$:
\begin{equation}
1 - e^{-\sigma_t \ell_{\max,t}} < \alpha_\mathrm{min} ,
\label{eq:skip}
\end{equation}
with $\ell_{\max,t}$ its longest edge. The kernels also skip any segment with $\alpha_i < \alpha_\mathrm{min}$, similar to how  3DGS~\cite{kerbl3Dgaussians} skips Gaussians below an opacity of $1/255$. This holds in training and at test time. It changes no geometry, only drops very faint segments from the image and from the gradients.

\paragraph{Gradients.} The density of a tetrahedron affects the contribution of its own segment and, through the transmittance, the contributions of all segments behind it.
Its gradient collects both.
The vertex features receive the gradient of the photometric loss, backpropagated through the MLP and weighted by their weights in the segment integrals.
The vertex positions receive gradients through the segment length and the barycentric coordinates of the entry and exit points.
The connectivity and the sort order are not differentiated, as they change only at retriangulation.
Forward and backward passes are custom CUDA kernels.

\paragraph{Densification.} In each view, a tetrahedron collects $\sum w_i e$ and $\sum w_i$ over the segments it contributes, with $w_i$ the blending weight of \cref{eq:alpha} and $e$ the error of the pixel.
A pixel counts toward its coverage when $w_i > 0.001$, so segments below the skip threshold add no error, as in rendering.

A tetrahedron becomes a candidate for splitting when, in at least one view, it covers at least 2 pixels and its weighted average error $\sum w_i e / \sum w_i$ is at least the average error over all pixels.
Among the candidates, the probability that a tetrahedron is split is proportional to its weighted error $\sum w_i e$ summed over all 40 views, so tetrahedra that cover many pixels or are seen in many views are split more often.
We draw the tetrahedra to split without replacement with the Gumbel top-$k$ trick~\cite{kool2019stochastic}.
When there are fewer candidates than vertices to add, all candidates are split and the step adds fewer vertices.

The depth mask uses the expected depth of the current model, rendered with the 40 views of the step.
It is the camera-space depth of the vertices, composited with the same weights as the features and divided by the accumulated opacity.
A segment is masked when it ends in front of 0.95 times this depth.
A segment that reaches beyond it is kept, so the tetrahedron that contains the surface still counts, not only those behind it.
Masked segments do not accumulate error, weight, or coverage, so a tetrahedron in front of the surface does not become a candidate through these pixels.
Pixels with accumulated opacity of at most 0.5 are not masked.
\Cref{sec:supp_dev} detail how we chose the factor 0.95, and \cref{tab:supp_depth} compares the rendered depth with monocular and multi-view depth maps for the mask.

\paragraph{Culling.} The ``With culling at 0.2'' and ``With culling at 0.1'' rows of \cref{tab:supp_abl_m360,tab:supp_abl_tnt} leave out a tetrahedron after each retriangulation when both its density and its opacity across its shortest edge $\ell_{\min,t}$ are below a threshold $\theta$,
\begin{equation}
\sigma_t < \theta \quad \text{and} \quad 1 - e^{-\sigma_t \ell_{\min,t}} < \theta .
\label{eq:cull}
\end{equation}
$\theta$ rises linearly from 0 at 4.5k iterations to 0.2 or 0.1 at 15k iterations.
The paper of Radiance Meshes~\cite{mai2026radiance} describes no culling, but its code culls from 4.5k at 0.1.
Culled tetrahedra stay in the triangulation and are re-evaluated at every retriangulation, so they can come back.
The culled set is fixed at 18k.

\paragraph{Retriangulation.} We use gDel3D's~\cite{cao2014gdel3d} GPU triangulation without its CPU star splaying, a final step that repairs the few faces that violate the Delaunay property.
Without splaying, fewer than 0.01\,\% of the faces violate it (\cref{tab:supp_splay}).
\Cref{tab:supp_splay} also gives how many faces of our final meshes the power-distance sort puts in the wrong visibility order.
Splaying runs on the CPU after every retriangulation, \ie every 10 iterations while the vertex positions are optimized.
On average over bicycle, garden and room, training is 1.17$\times$ faster and PSNR is 0.02\,dB lower without splaying.

\begin{table}[H]
\centering
\caption{Faces that violate the Delaunay property after retriangulating the final vertices without star splaying, and the training speedup without it.
Wrong order is the share of the interior faces of the final mesh, which also contains the position step after the last retriangulation, whose two tetrahedra the power-distance sort orders wrongly.
This happens from every camera exactly at the faces that are not locally Delaunay, since the circumspheres of two tetrahedra that share a face have equal power distance only in the plane of that face.}
\label{tab:supp_splay}
\footnotesize
\setlength{\tabcolsep}{2.3pt}
\begin{tabular}{@{}l ccccc@{}}
\toprule
 & Interior faces & Violating faces & Share & Wrong order & Speedup \\
\midrule
Bicycle & 65.1M & 4{,}723 & 0.0073\,\% & 0.057\,\% & 1.25$\times$ \\
Garden & 65.1M & 4{,}170 & 0.0064\,\% & 0.059\,\% & 1.16$\times$ \\
Room & 39.2M & 2{,}406 & 0.0061\,\% & 0.060\,\% & 1.11$\times$ \\
\bottomrule
\end{tabular}
\end{table}

\PAR{Deferred decoding.}
Deferred Neural Rendering~\cite{thies2019deferred} rasterizes learned neural textures and decodes them per pixel using a  network.
SNeRG~\cite{hedman2021baking} and MERF~\cite{reiser2023merf} composite colors and features along each ray through a baked grid and decode the view-dependent part of the color once per pixel with a small MLP.
MobileNeRF~\cite{chen2023mobilenerf} decodes the features of rasterized polygons per pixel in the fragment shader, and Ref-GS~\cite{zhang2024ref} shades rasterized Gaussians once per pixel.
We decode in the same way, mainly so that the MLP does not run in the rasterization kernel.
We composite the features of all segments along a ray into one feature vector $F$ (\cref{eq:sum}) and evaluate the color MLP once per pixel on $F$ and the encoded view direction (\cref{eq:direnc}), instead of once per segment.

\section{Experimental setup}
\label{sec:supp_setup}

\subsection{Datasets}
\label{sec:supp_data}

We use the images and train/test splits as NerfBaselines~\cite{kulhanek2024nerfbaselines} provides them.
The Mip-NeRF~360 outdoor scenes (bicycle, flowers, garden, stump and treehill) are downsampled $4\times$, to about $1240\times820$ pixels, and the indoor scenes (bonsai, counter, kitchen and room) $2\times$, to about $1560\times1040$.
This is the protocol of 3DGS~\cite{kerbl3Dgaussians}, Radiant Foam~\cite{govindarajan2025radfoam} and Radiance Meshes~\cite{mai2026radiance}.
The Tanks and Temples~\cite{knapitsch2017tanks} images are downsampled $2\times$, to about $980\times545$.
In both datasets every eighth image is a test view.
The 6 indoor scenes of Tanks and Temples are auditorium, ballroom, church, courtroom, meetingroom and museum, and the other 15 are outdoor scenes.
As in NerfBaselines, \cref{tab:nvs} reports LPIPS with the AlexNet network.

\subsection{Baseline configurations}
\label{sec:supp_baselines}

\begin{table*}[!tp]
\centering
\caption{Baselines on Tanks and Temples with each configuration we ran, the default configuration of the code and the Mip-NeRF~360 configuration of the matching scene type.
Bold marks the configuration used in \cref{tab:nvs}, the one with the better mean PSNR.
A dash marks a run we did not do.
We run each Mip-NeRF~360 configuration only on its matching scene type of Tanks and Temples.
The indoor configuration of PowerFoam~\cite{govindarajan2026powerfoam} and the outdoor configuration of VoroTracing~\cite{taveira2026vorotracing} are their defaults, so they have no separate row.
For ours, 5M is the vertex count of the outdoor configuration, and 3M that of the default configuration.
The column min gives the training time in minutes.
\rfmark{}Train with our triangulation.}
\footnotesize
\setlength{\tabcolsep}{3pt}
\begin{tabular}{@{}l l cccc cccc@{}}
\toprule
 & & \multicolumn{4}{c}{T\&T indoor (6)} & \multicolumn{4}{c}{T\&T outdoor (15)} \\
\cmidrule(lr){3-6}\cmidrule(lr){7-10}
Method & Configuration & PSNR & SSIM & LPIPS & min & PSNR & SSIM & LPIPS & min \\
\midrule
Radiance Meshes & single & \textbf{21.69} & \textbf{0.794} & \textbf{0.182} & \textbf{271} & \textbf{23.75} & \textbf{0.833} & \textbf{0.159} & \textbf{325} \\
\midrule
PowerFoam & default (= M360 indoor) & \textbf{21.52} & \textbf{0.727} & \textbf{0.276} & \textbf{33} & \textbf{23.14} & \textbf{0.768} & \textbf{0.227} & \textbf{36} \\
 & M360 outdoor & -- & -- & -- & -- & 22.78 & 0.756 & 0.239 & 60 \\
\midrule
Radiant Foam & default & 20.77 & 0.764 & 0.190 & 170 & \textbf{22.71}\rfmark{} & \textbf{0.797}\rfmark{} & \textbf{0.182}\rfmark{} & \textbf{146}\rfmark{} \\
 & M360 indoor & \textbf{20.80} & \textbf{0.764} & \textbf{0.189} & \textbf{176} & -- & -- & -- & -- \\
 & M360 outdoor & -- & -- & -- & -- & 21.93 & 0.790 & 0.186 & 263 \\
\midrule
VoroTracing & default (= M360 outdoor) & 21.33 & 0.754 & 0.221 & 81 & \textbf{23.56} & \textbf{0.802} & \textbf{0.169} & \textbf{77} \\
 & M360 indoor & \textbf{21.46} & \textbf{0.756} & \textbf{0.218} & \textbf{84} & -- & -- & -- & -- \\
\midrule
Ours & default & \textbf{23.04} & \textbf{0.817} & \textbf{0.136} & \textbf{38} & \textbf{25.06} & \textbf{0.846} & \textbf{0.121} & \textbf{37} \\
 & outdoor, 3M & -- & -- & -- & -- & 24.62 & 0.833 & 0.146 & 31 \\
 & outdoor, 5M & -- & -- & -- & -- & 24.59 & 0.834 & 0.142 & 43 \\
\bottomrule
\end{tabular}
\label{tab:supp_baselines}
\end{table*}

\Cref{tab:supp_baselines} lists the runs of the baselines on Tanks and Temples. The indoor configurations of Radiant Foam and VoroTracing help a little on the indoor scenes. The outdoor configurations of PowerFoam~\cite{govindarajan2026powerfoam} and Radiant Foam~\cite{govindarajan2025radfoam} are 0.4--0.8\,dB worse than their defaults on the outdoor scenes.
Ours behaves the same way, as its outdoor configuration costs 0.44\,dB there with 3M vertices and 0.47\,dB with 5M.

The default configurations of PowerFoam and Radiant Foam have fewer points than their outdoor ones, 500k instead of 1.2M and 2.1M instead of 4.2M, similar to our default configuration that has 3M instead of 5M vertices. The indoor and outdoor configurations of VoroTracing have the same number of points and differ only in their regularization.

\subsection{Reproducing the baselines}
\label{sec:supp_repro}

\begin{table*}[!tp]
\centering
\caption{Mip-NeRF~360 test PSNR of the baselines as reported in their papers, trained with their official code (run by us without NerfBaselines, one run per scene on the same data and GPU), and trained in a wrapper inside NerfBaselines as in all our other tables. ``Own PSNR'' is the PSNR computed by the official code of PowerFoam, Radiant Foam and VoroTracing. Means are over the scenes the paper reports (PowerFoam, Radiant Foam and VoroTracing do not report results for flowers and treehill). Radiance Meshes reports only the outdoor and indoor means, and its authors' per-scene results are released with its code. $^b$The official code renders one test view blank, and the other views average 21.52.}
\label{tab:supp_repro}
\scriptsize
\setlength{\tabcolsep}{2.4pt}
\begin{tabular}{@{}l l ccccc cccc ccc@{}}
\toprule
 & & \multicolumn{5}{c}{Outdoor} & \multicolumn{4}{c}{Indoor} & \multicolumn{3}{c}{Mean} \\
\cmidrule(lr){3-7}\cmidrule(lr){8-11}\cmidrule(l){12-14}
Method & & bicycle & flowers & garden & stump & treehill & room & counter & kitchen & bonsai & Out. & In. & All \\
\midrule
Radiance Meshes & Paper (means only) & -- & -- & -- & -- & -- & -- & -- & -- & -- & 24.38 & 30.61 & 27.15 \\
 & Authors' per-scene results & 25.06 & 21.56 & 27.04 & 26.58 & 22.97 & 30.87 & 28.60 & 30.76 & 30.99 & 24.64 & 30.31 & 27.16 \\
 & Official code & 25.02 & 21.50 & 26.97 & 26.42 & 23.03 & 30.58 & 28.58 & 30.81 & 30.93 & 24.59 & 30.22 & 27.09 \\
 & NerfBaselines & 24.98 & 21.61 & 26.99 & 26.39 & 22.97 & 30.58 & 28.56 & 30.84 & 30.62 & 24.59 & 30.15 & 27.06 \\
\cmidrule(l){2-14}
 & NerfBaselines $-$ paper &  &  &  &  &  &  &  &  &  & +0.21 & $-$0.46 & $-$0.09 \\
 & NerfBaselines $-$ authors & $-$0.08 & +0.05 & $-$0.05 & $-$0.19 & 0.00 & $-$0.29 & $-$0.05 & +0.08 & $-$0.38 & $-$0.05 & $-$0.16 & $-$0.10 \\
 & NerfBaselines $-$ official & $-$0.04 & +0.11 & +0.02 & $-$0.03 & $-$0.06 & 0.00 & $-$0.02 & +0.03 & $-$0.31 & 0.00 & $-$0.08 & $-$0.03 \\
\midrule
PowerFoam & Paper & 24.18 & -- & 27.10 & 25.35 & -- & 31.23 & 29.81 & 31.42 & 33.32 & 25.54 & 31.45 & 28.92 \\
 & Official code, own PSNR & 24.10 & 21.32 & 26.94 & 25.22 & 22.36 & 31.00 & 29.67 & 31.35 & 33.24 & 25.42 & 31.31 & 28.79 \\
 & Official code & 24.09 & 21.30 & 26.94 & 25.13 & 22.33 & 30.95 & 29.65 & 31.32 & 33.21 & 25.39 & 31.28 & 28.75 \\
 & NerfBaselines & 24.07 & 21.31 & 26.96 & 25.11 & 22.36 & 31.08 & 29.67 & 31.09 & 33.16 & 25.38 & 31.25 & 28.73 \\
\cmidrule(l){2-14}
 & Official own PSNR $-$ paper & $-$0.08 &  & $-$0.16 & $-$0.13 &  & $-$0.23 & $-$0.14 & $-$0.07 & $-$0.08 & $-$0.12 & $-$0.13 & $-$0.13 \\
 & NerfBaselines $-$ paper & $-$0.11 &  & $-$0.14 & $-$0.24 &  & $-$0.15 & $-$0.14 & $-$0.33 & $-$0.16 & $-$0.16 & $-$0.20 & $-$0.18 \\
 & NerfBaselines $-$ official & $-$0.02 & +0.01 & +0.03 & $-$0.02 & +0.03 & +0.13 & +0.02 & $-$0.23 & $-$0.05 & 0.00 & $-$0.03 & $-$0.02 \\
\midrule
Radiant Foam & Paper & 24.19 & -- & 26.58 & 25.48 & -- & 30.87 & 28.59 & 31.40 & 32.15 & 25.42 & 30.75 & 28.47 \\
 & Official code, own PSNR & 24.26 & 21.43 & 26.55 & 25.49 & 22.14 & 30.83 & 28.48 & 31.25 & 32.22 & 25.43 & 30.70 & 28.44 \\
 & Official code & 24.25 & 21.41 & 26.55 & 25.40 & 22.12 & 30.80 & 28.46 & 31.20 & 32.17 & 25.40 & 30.66 & 28.41 \\
 & NerfBaselines & 24.14 & 21.35 & 26.56 & 25.33 & 22.26 & 30.75 & 28.53 & 31.22 & 32.20 & 25.34 & 30.67 & 28.39 \\
\cmidrule(l){2-14}
 & Official own PSNR $-$ paper & +0.07 &  & $-$0.03 & +0.01 &  & $-$0.04 & $-$0.11 & $-$0.15 & +0.07 & +0.02 & $-$0.06 & $-$0.02 \\
 & NerfBaselines $-$ paper & $-$0.05 &  & $-$0.02 & $-$0.15 &  & $-$0.12 & $-$0.06 & $-$0.18 & +0.05 & $-$0.07 & $-$0.08 & $-$0.08 \\
 & NerfBaselines $-$ official & $-$0.11 & $-$0.06 & +0.02 & $-$0.07 & +0.14 & $-$0.06 & +0.07 & +0.02 & +0.02 & $-$0.06 & +0.01 & $-$0.02 \\
\midrule
VoroTracing & Paper & 24.29 & -- & 26.98 & 25.94 & -- & 31.23 & 29.48 & 31.54 & 33.42 & 25.74 & 31.42 & 28.98 \\
 & Official code, own PSNR & 24.36 & 20.72$^b$ & 27.00 & 26.04 & 22.56 & 31.45 & 29.48 & 31.64 & 33.13 & 25.80 & 31.43 & 29.01 \\
 & Official code & 24.35 & 20.69$^b$ & 26.99 & 25.95 & 22.52 & 31.39 & 29.46 & 31.61 & 33.10 & 25.76 & 31.39 & 28.98 \\
 & NerfBaselines & 24.34 & 21.54 & 27.00 & 25.95 & 22.59 & 31.32 & 29.43 & 31.54 & 33.13 & 25.76 & 31.35 & 28.96 \\
\cmidrule(l){2-14}
 & Official own PSNR $-$ paper & +0.07 &  & +0.02 & +0.10 &  & +0.22 & 0.00 & +0.10 & $-$0.29 & +0.06 & +0.01 & +0.03 \\
 & NerfBaselines $-$ paper & +0.05 &  & +0.02 & +0.01 &  & +0.09 & $-$0.05 & 0.00 & $-$0.29 & +0.03 & $-$0.06 & $-$0.02 \\
 & NerfBaselines $-$ official & 0.00 & +0.85$^b$ & +0.01 & 0.00 & +0.07 & $-$0.07 & $-$0.03 & $-$0.07 & +0.03 & 0.00 & $-$0.04 & $-$0.02 \\
\bottomrule
\end{tabular}
\end{table*}

\begin{table*}[!tp]
\centering
\caption{Scene-type settings. Each row swaps one setting relative to what the scene type uses; differences to ours (A dash marks the row with the scene type's own config). Other encoding: frequency (four frequencies from 1 to 2 cycles) on M360 outdoor, SH degree 3 on all other scenes, same feature dim. With depth loss: Depth Anything 3 (DA3)~\cite{depthanything3} depth loss with weight 0.1 until 5k. Pos.\ lr $\times2$: off on M360 outdoor, on elsewhere. Default config.: our default configuration, with its 3M vertices, on Mip-NeRF~360 outdoor, so one configuration for all scenes. Default config., 5M: the same with the 5M vertices of the outdoor configuration. Outdoor config., 3M: the settings of the outdoor configuration with the 3M vertices of the other scenes.}
\label{tab:supp_scene}
\footnotesize
\setlength{\tabcolsep}{2.5pt}
\begin{tabular}{@{}l ccc ccc ccc@{}}
\toprule
 & \multicolumn{3}{c}{M360 outdoor (5)} & \multicolumn{3}{c}{Indoor (10)} & \multicolumn{3}{c}{T\&T outdoor (15)} \\
\cmidrule(lr){2-4}\cmidrule(lr){5-7}\cmidrule(lr){8-10}
Method & PSNR & SSIM & LPIPS & PSNR & SSIM & LPIPS & PSNR & SSIM & LPIPS \\
\midrule
Ours & 24.42 & 0.718 & 0.188 & 26.72 & 0.862 & 0.124 & 25.06 & 0.846 & 0.121 \\
\midrule
Other encoding & $-$0.22 & $-$0.010 & +0.009 & $-$0.12 & $-$0.002 & +0.003 & $-$0.03 & $-$0.001 & +0.002 \\
With depth loss & 0.00 & +0.001 & $-$0.002 & $-$0.12 & 0.000 & $-$0.001 & $-$0.23 & $-$0.002 & 0.000 \\
Pos.\ lr $\times2$ toggled & $-$0.07 & $-$0.001 & $-$0.004 & $-$0.02 & 0.000 & +0.002 & $-$0.04 & $-$0.002 & +0.005 \\
Default config. & $-$1.04 & $-$0.034 & +0.009 & -- & -- & -- & -- & -- & -- \\
Default config., 5M & $-$0.87 & $-$0.027 & $-$0.003 & -- & -- & -- & -- & -- & -- \\
Outdoor config., 3M & -- & -- & -- & $-$0.63 & $-$0.011 & +0.022 & $-$0.44 & $-$0.013 & +0.025 \\
\bottomrule
\end{tabular}
\end{table*}

\Cref{tab:supp_repro} checks the NerfBaselines runs of the baselines, which all our tables use, on Mip-NeRF~360 against their official code and their papers.
All rows use the standard PSNR of NerfBaselines, computed from the mean squared error over all three color channels.
The official code of PowerFoam, Radiant Foam and VoroTracing instead averages the PSNRs of the R, G and B channels, which we call their ``own PSNR''.
Since the logarithm is concave, this is never lower than the standard PSNR, and on these runs it is 0.03--0.04\,dB higher on average.
On average, the NerfBaselines runs are within 0.03\,dB of the official code, and within 0.15\,dB on every scene except Radiance Meshes bonsai ($-0.31$\,dB), PowerFoam kitchen ($-0.23$\,dB), and VoroTracing flowers, where the official code renders one test view entirely white (\cref{tab:supp_repro}).
Over the scenes each paper reports, the NerfBaselines runs are within 0.1\,dB of the published results, except PowerFoam ($-0.18$\,dB), whose official code is already 0.13\,dB below its paper when evaluated with its own PSNR.
Radiance Meshes reports only outdoor and indoor means.
The NerfBaselines reruns are 0.21\,dB above and 0.46\,dB below them, but its per-scene results released with the code differ from them by $+0.26$ and $-0.30$\,dB as well.

\subsection{Development and constants}
\label{sec:supp_dev}
We developed the method on bicycle, garden and room of Mip-NeRF~360, mostly at reduced resolution with 10k iterations.
We found $\alpha_\mathrm{min}$ using a sweep on the garden, bicycle, and room scenes. The default 3DGS~\cite{kerbl3Dgaussians} threshold of $1/255$ was too strong, and $\alpha_\mathrm{min} = 0.001$ was the largest threshold that essentially costs no quality.
The densification ignores the error in front of 0.95 times the expected depth (\cref{sec:optimization}).
We chose this factor on bicycle at reduced resolution, where factors from 0.8 to 0.99 change test PSNR by at most 0.11\,dB, while dropping the depth mask costs 0.35\,dB.
Each densification step renders 40 random training views.
Radiance Meshes~\cite{mai2026radiance} renders 200, or all of them when there are fewer.
On bicycle, using all 169 training views, as Radiance Meshes does there, was not better.
A tetrahedron must cover at least 2 pixels in a view.
Radiance Meshes requires at least 9 pixels, and we found 2 to be enough.

As Radiance Meshes, we densify every 500 iterations from 2k iterations and fix the mesh at 18k iterations.
Radiance Meshes adds vertices where its error statistics exceed thresholds, so its vertex count follows no fixed schedule, and it has no exponents $p_k$ or target counts.
We only approximate the vertex counts of its runs with our schedule.
For a run with $n_0$ initial and $n_\mathrm{end}$ final vertices and $n_k$ vertices at our step $k$, we set $p_k = \log(n_k / n_0) / \log(n_\mathrm{end} / n_0)$, so our target count follows the same curve, scaled to our budget.
The fast schedule takes these exponents from its bicycle run, which reaches its final count at 7.5k iterations, and the slow schedule takes their mean over its garden and room runs, which reach it at 15.5k iterations.
Both gave the same quality on the 9 Mip-NeRF~360 scenes, so we use the fast one.
The exponents $p_k$ of the faster schedule at its 12 steps are 0.30, 0.55, 0.70, 0.80, 0.87, 0.91, 0.93, 0.95, 0.97, 0.98, 0.99 and 1 (\cref{sec:optimization}).
This schedule is not elegant.
The thresholds of Radiance Meshes are absolute values on its own error statistics and do not transfer to our method, but we wanted our mesh to grow like that of Radiance Meshes, since its densification works well.
The stored exponents could be replaced by a simpler parametrization.
For example, $p_k = 1 - 0.70\, e^{-k / 2.44}$, with $p_k = 1$ at the last step, matches the fast schedule within 0.01.
Written in iterations instead of steps, such a form would also give a longer or shorter densification period by scaling its time constant and moving its last step.

\subsection{Scene-type settings}
\label{sec:supp_scene_types}

Changing a single setting changes the results by at most about 0.2\,dB, but the configuration of one scene type does not transfer to another (\cref{tab:supp_scene}).
SH directions suit Mip-NeRF~360 outdoor, and the Fourier features suit the other scenes.
The doubled position learning rate helps only on Mip-NeRF~360 outdoor.
Our default configuration loses about 1\,dB on Mip-NeRF~360 outdoor, and most of this loss remains with the 5M vertices of the outdoor configuration.
So the loss comes from the settings and not from the vertex budget.
As no single setting loses more than about 0.2\,dB, it comes from several settings together.
The outdoor configuration is worse than the default one on the other scene types.
The same holds for PowerFoam and Radiant Foam, whose Mip-NeRF~360 outdoor configurations are worse than their defaults there (\cref{sec:supp_baselines}).
We therefore use our default configuration on all Tanks and Temples scenes.

\paragraph{Choices behind the two configurations.}
Starting from the 64 feature dimensions of Tetra-NeRF~\cite{kulhanek2023tetranerf}, we decreased them to 32 in the default configuration and 16 in the outdoor one, which makes training faster.
In an earlier configuration, 48 dimensions were 0.1--0.2\,dB better than 32 on the default scenes, but training took 15--25\,\% longer.
Tetra-NeRF trains with an L2 loss.
Following common practice, we switched to L1, which was 0.3\,dB better on garden in an early version of our method, and then added an SSIM~\cite{wang2004ssim} term (\cref{eq:loss}) with weight $\lambda_\mathrm{SSIM} = 0.2$, as 3DGS~\cite{kerbl3Dgaussians}, MeshSplatting~\cite{Held2025MeshSplatting} and Radiance Meshes~\cite{mai2026radiance}.
The SSIM term improved the indoor scenes, but not the outdoor ones, where the full weight cost 0.09\,dB PSNR in an earlier configuration and only slightly improved SSIM and LPIPS~\cite{zhang2018lpips}.
So the outdoor configuration uses $\lambda_\mathrm{SSIM} = 0.1$.

\subsection{Hyperparameters}
\label{sec:supp_hparams}

\Cref{tab:supp_hparams} lists the hyperparameters shared by both configurations, and \cref{tab:supp_hparams_cfg} those that differ.
Many of them are simply those of Radiance Meshes~\cite{mai2026radiance}, as marked in the tables.
We optimize all parameters with Adam~\cite{kingma2015adam} with $\epsilon = 10^{-15}$, as Radiance Meshes.
The learning rates decay log-linearly from their initial to their final value until the mesh is fixed at 18k and stay constant after.
As in Radiance Meshes, the position learning rate is multiplied by the scene scale, the largest offset of a training camera from their mean position.
Every vertex starts with the log density $-4$, \ie with the density $e^{-4} \approx 0.02$, so the initial mesh is nearly transparent.
From 18k, when the mesh is fixed, each tetrahedron gets its own log density, initialized with the log of its density $\sigma_t$ from its vertices (\cref{eq:density}) and then optimized directly, while the vertex densities are no longer trained.
These are the per-tetrahedron densities of the main paper (\cref{sec:supp_density}).
The position gradients are accumulated over the 10 iterations between two retriangulations and applied in one step, and the mesh is retriangulated afterwards.
The color MLP uses ReLU activations and a sigmoid output and runs in bf16.
The Fourier direction encoding uses four frequencies from 1 to 2 cycles (\cref{eq:direnc}).
We render one ray per pixel, and supersampling with $2 \times 2$ rays per pixel is an option at test time only.
The exponents of the densification schedule are given in \cref{sec:supp_dev}.

\begin{table}[H]
\centering
\caption{Hyperparameters shared by both configurations.}
\label{tab:supp_hparams}
\scriptsize
\setlength{\tabcolsep}{3pt}
\begin{tabular}{@{}l l@{}}
\toprule
Hyperparameter & Value \\
\midrule
Iterations & 30k (as \cite{mai2026radiance}) \\
Mesh fixed & at 18k (as \cite{mai2026radiance}) \\
Retriangulation & every 10 iterations (as \cite{mai2026radiance}) \\
Outer-sphere vertices & 1000 (as \cite{mai2026radiance}) \\
\midrule
Vertex log densities & init $-4$ (as \cite{mai2026radiance}), lr 0.1 until 18k \\
Per-tetrahedron log densities & from 18k, init from the vertices, lr 0.01 \\
Features & init $\mathcal{N}(0, 0.1^2)$, initial lr $3 \cdot 10^{-2}$ \\
Color MLP & 2 hidden layers of 256, lr $10^{-3} \to 10^{-4}$ \\
\midrule
Faint-segment regularizer & $\theta_\mathrm{f} = 0.2$, $\lambda_\mathrm{f} = 3 \cdot 10^{-4}$, from 10k \\
Skip threshold & $\alpha_\mathrm{min} = 0.001$ \\
\midrule
Densification steps & every 500 from 2k (as \cite{mai2026radiance}) to 7.5k \\
Views per step & 40 \\
Depth-mask factor & 0.95 \\
\bottomrule
\end{tabular}
\end{table}

\begin{table}[H]
\centering
\caption{Hyperparameters that differ between the default configuration and the outdoor configuration (Mip-NeRF~360 outdoor), as in \cref{tab:settings}.}
\label{tab:supp_hparams_cfg}
\scriptsize
\setlength{\tabcolsep}{3pt}
\begin{tabular}{@{}l cc@{}}
\toprule
Hyperparameter & Default & Outdoor \\
\midrule
Vertex budget & 3M & 5M \\
Feature dimension & 32 & 16 \\
Direction encoding & Fourier & SH, degree 3 \\
Position lr & $10^{-4} \to 10^{-6}$ (as \cite{mai2026radiance}) & $2 \cdot 10^{-4} \to 2 \cdot 10^{-6}$ \\
Final feature lr & $3 \cdot 10^{-3}$ & $3 \cdot 10^{-4}$ \\
$\lambda_\mathrm{SSIM}$ & 0.2 (as \cite{mai2026radiance}) & 0.1 \\
\bottomrule
\end{tabular}
\end{table}

\section{Additional results}
\label{sec:supp_results}

\subsection{Results per scene type}
\label{sec:supp_types}

\begin{table*}[!tp]
\centering
\caption{\Cref{tab:nvs} on Mip-NeRF~360, split into outdoor and indoor scenes. It extends \cref{tab:nvs} with LPIPS from the VGG network next to the AlexNet one and with the training time per scene type, and leaves out Radiance Meshes without culling. The column min gives the training time in minutes. Colors as in \cref{tab:nvs}.}
\label{tab:supp_m360}
\footnotesize
\setlength{\tabcolsep}{2.5pt}
\begin{tabular}{@{}l ccccc ccccc@{}}
\toprule
 & \multicolumn{5}{c}{Mip-NeRF~360 outdoor (5)} & \multicolumn{5}{c}{Mip-NeRF~360 indoor (4)} \\
\cmidrule(lr){2-6}\cmidrule(lr){7-11}
 & & & \multicolumn{2}{c}{LPIPS} & & & & \multicolumn{2}{c}{LPIPS} & \\
\cmidrule(lr){4-5}\cmidrule(lr){9-10}
Method & PSNR & SSIM & Alex & VGG & min & PSNR & SSIM & Alex & VGG & min \\
\midrule
Radiance Meshes          & \fst{24.59} & \fst{0.726} & \snd{0.203} & \snd{0.290} & 335 & 30.15 & \snd{0.918} & 0.130 & \trd{0.241} & 353 \\
PowerFoam                & 23.96 & 0.669 & 0.222 & 0.325 & \snd{54} & \trd{31.25} & 0.912 & \trd{0.122} & \trd{0.241} & \fst{48} \\
Radiant Foam             & 23.93 & 0.669 & 0.256 & 0.351 & 165 & 30.67 & 0.901 & 0.127 & 0.261 & 103 \\
VoroTracing              & \trd{24.28} & \trd{0.695} & \trd{0.209} & \trd{0.298} & \trd{72} & \snd{31.35} & \trd{0.915} & \snd{0.112} & \snd{0.216} & \snd{59} \\
\midrule
Ours                     & \snd{24.42} & \snd{0.718} & \fst{0.188} & \fst{0.271} & \fst{50} & \fst{32.24} & \fst{0.930} & \fst{0.106} & \fst{0.204} & \trd{63} \\
Ours + SS                & \dfst{24.62} & \dsnd{0.724} & \dsnd{0.200} & \dfst{0.266} & \dsnd{50} & \dfst{32.31} & \dfst{0.932} & \dsnd{0.110} & \dsnd{0.205} & 63 \\
Ours, RM \#vert.         & 24.08 & 0.693 & 0.224 & 0.318 & \dfst{28} & 32.05 & 0.927 & 0.114 & 0.221 & \dfst{42} \\
Ours, RM \#vert. + SS    & 24.24 & 0.698 & 0.236 & 0.314 & \dfst{28} & \dtrd{32.12} & \dtrd{0.928} & 0.118 & 0.221 & \dfst{42} \\
\bottomrule
\end{tabular}
\end{table*}

\begin{table*}[!tp]
\centering
\caption{\Cref{tab:nvs} on Tanks and Temples, split into outdoor and indoor scenes, which \cref{tab:nvs} reports together. It extends \cref{tab:nvs} with LPIPS from the VGG network next to the AlexNet one and with the training time per scene type, and leaves out Radiance Meshes without culling. The column min gives the training time in minutes. Colors as in \cref{tab:nvs}. \rfmark{}Train with our triangulation.}
\label{tab:supp_tnt}
\footnotesize
\setlength{\tabcolsep}{2.5pt}
\begin{tabular}{@{}l ccccc ccccc@{}}
\toprule
 & \multicolumn{5}{c}{Tanks and Temples outdoor (15)} & \multicolumn{5}{c}{Tanks and Temples indoor (6)} \\
\cmidrule(lr){2-6}\cmidrule(lr){7-11}
 & & & \multicolumn{2}{c}{LPIPS} & & & & \multicolumn{2}{c}{LPIPS} & \\
\cmidrule(lr){4-5}\cmidrule(lr){9-10}
Method & PSNR & SSIM & Alex & VGG & min & PSNR & SSIM & Alex & VGG & min \\
\midrule
Radiance Meshes          & \snd{23.75} & \snd{0.833} & \snd{0.159} & \snd{0.252} & 325 & \snd{21.69} & \snd{0.794} & \snd{0.182} & \snd{0.286} & 271 \\
PowerFoam                & 23.14 & 0.768 & 0.227 & 0.331 & \fst{36} & \trd{21.52} & 0.727 & 0.276 & 0.385 & \fst{33} \\
Radiant Foam             & 22.71\rfmark{} & 0.797\rfmark{} & 0.182\rfmark{} & 0.292\rfmark{} & 146\rfmark{} & 20.80 & \trd{0.764} & \trd{0.189} & \trd{0.310} & 176 \\
VoroTracing              & \trd{23.56} & \trd{0.802} & \trd{0.169} & \trd{0.277} & \trd{77} & 21.46 & 0.756 & 0.218 & 0.330 & \trd{84} \\
\midrule
Ours                     & \fst{25.06} & \fst{0.846} & \fst{0.121} & \fst{0.209} & \snd{37} & \fst{23.04} & \fst{0.817} & \fst{0.136} & \fst{0.241} & \snd{38} \\
Ours + SS                & \dfst{25.19} & \dfst{0.850} & \dsnd{0.131} & \dfst{0.209} & \dtrd{37} & \dsnd{23.12} & \dfst{0.821} & \dsnd{0.142} & \dfst{0.237} & \dtrd{38} \\
Ours, RM \#vert.         & 25.04 & \dtrd{0.843} & \dtrd{0.135} & \dsnd{0.229} & \dfst{22} & \dtrd{23.11} & \dtrd{0.817} & \dtrd{0.145} & 0.255 & \dfst{23} \\
Ours, RM \#vert. + SS    & \dsnd{25.16} & \dsnd{0.846} & 0.146 & \dsnd{0.229} & \dfst{22} & \dfst{23.19} & \dsnd{0.820} & 0.152 & \dtrd{0.253} & \dfst{23} \\
\bottomrule
\end{tabular}
\end{table*}

\Cref{tab:supp_m360,tab:supp_tnt} split \cref{tab:nvs} by scene type and add LPIPS with the VGG network.
\Cref{tab:supp_scenes_m360,tab:supp_scenes_tnt}, at the end of the supplementary material, list the results on every scene.
NerfBaselines evaluates Mip-NeRF~360 with its NeRF evaluation protocol, which includes LPIPS-VGG, and Tanks and Temples with its default protocol, which does not.
On Tanks and Temples, we therefore compute LPIPS-VGG from the same test renders with NerfBaselines' metric code.
On Tanks and Temples we gain about 1.3\,dB PSNR over Radiance Meshes on both scene types.
We are best in LPIPS-VGG on all four scene types, as in LPIPS-Alex.
Supersampling makes LPIPS-Alex slightly worse, while LPIPS-VGG stays about the same.
The lowest crops of the baselines in \cref{fig:qual} are local windows, and on full Courtroom test images the baselines reach 19.8--20.8\,dB.
To test whether the baselines lose on Tanks and Temples through global color or brightness shifts, we fit each render to its ground truth with a per-image affine color correction before computing PSNR.
This raises every method by 0.5--1.1\,dB on Tanks and Temples, and our margin over the best baseline stays 1.4\,dB (25.14 against 23.71\,dB).

\subsection{Comparison with other methods}
\label{sec:supp_others}

\begin{table*}[!tp]
\centering
\caption{Comparison with NeRF and Gaussian splatting methods and with the baselines of \cref{tab:nvs}, sorted by Mip-NeRF~360 PSNR.
It extends \cref{tab:tetranerf} with more NeRF and Gaussian splatting methods, the baselines and Tanks and Temples, and leaves out the variants of our method on the Tetra-NeRF points.
The results of the NeRF and Gaussian splatting methods are those published by NerfBaselines~\cite{kulhanek2024nerfbaselines}, trained on the same A100-40GB GPU type, with the same test images and LPIPS network as ours.
Tetra-NeRF is our NerfBaselines run of \cref{tab:tetranerf}, and the baselines are our runs of \cref{tab:nvs}.
The rows of ours and of these baselines are shaded.
We highlight the \colorbox{rkfirst}{first}, \colorbox{rksecond}{second} and \colorbox{rkthird}{third} best result in each column.
A dash marks results that NerfBaselines does not publish.
The columns Train give the training time in minutes.
\rfmark{}Train with our triangulation.}
\label{tab:supp_others}
\footnotesize
\setlength{\tabcolsep}{3pt}
\definecolor{rowhl}{gray}{0.92}
\begin{tabular}{@{}l c ccc cc c ccc c@{}}
\toprule
 & & \multicolumn{3}{c}{Mip-NeRF~360 (9)} & Outdoor (5) & Indoor (4) & Train & \multicolumn{3}{c}{Tanks and Temples (21)} & Train \\
\cmidrule(lr){3-5}\cmidrule(lr){6-6}\cmidrule(lr){7-7}\cmidrule(lr){8-8}\cmidrule(lr){9-11}\cmidrule(l){12-12}
Method & Iter. & PSNR & SSIM & LPIPS & PSNR & PSNR & min & PSNR & SSIM & LPIPS & min \\
\midrule
Zip-NeRF~\cite{barron2023zipnerf} & 200k & \fst{28.55} & \snd{0.829} & \fst{0.138} & \fst{25.57} & \fst{32.28} & 330 & \fst{24.63} & \fst{0.840} & \snd{0.131} & 344 \\
3DGS-MCMC~\cite{kheradmand2024mcmc} & 30k & \snd{27.98} & \fst{0.835} & \snd{0.148} & \snd{25.18} & 31.49 & 41 & -- & -- & -- & -- \\
\rowcolor{rowhl}\textbf{Ours} & \textbf{30k} & \trd{\textbf{27.90}} & \textbf{0.812} & \trd{\textbf{0.152}} & \textbf{24.42} & \snd{\textbf{32.24}} & \textbf{56} & \snd{\textbf{24.48}} & \snd{\textbf{0.838}} & \fst{\textbf{0.125}} & \textbf{37} \\
Mip-NeRF~360~\cite{barron2022mipnerf360} & 250k & 27.68 & 0.792 & 0.195 & 24.48 & \trd{31.68} & 1815 & -- & -- & -- & -- \\
Mip-Splatting~\cite{Yu2023MipSplatting} & 30k & 27.49 & \trd{0.815} & 0.183 & \trd{24.73} & 30.95 & 26 & \trd{23.93} & \trd{0.833} & 0.166 & \trd{16} \\
3DGS~\cite{kerbl3Dgaussians} & 30k & 27.43 & 0.814 & 0.180 & 24.64 & 30.93 & \trd{23} & 23.83 & 0.831 & \trd{0.165} & \snd{14} \\
\rowcolor{rowhl}VoroTracing~\cite{taveira2026vorotracing} & 20k & 27.43 & 0.793 & 0.166 & 24.28 & 31.35 & 66 & 22.96 & 0.789 & 0.183 & 79 \\
Taming 3DGS~\cite{mallick2024taming} & 30k & 27.22 & 0.793 & 0.234 & 24.45 & 30.68 & \snd{6} & -- & -- & -- & -- \\
\rowcolor{rowhl}PowerFoam~\cite{govindarajan2026powerfoam} & 30k & 27.20 & 0.777 & 0.178 & 23.96 & 31.25 & 51 & 22.68 & 0.756 & 0.241 & 35 \\
\rowcolor{rowhl}Radiance Meshes~\cite{mai2026radiance} & 30k & 27.06 & 0.811 & 0.171 & 24.59 & 30.15 & 343 & 23.16 & 0.822 & 0.166 & 310 \\
\rowcolor{rowhl}Radiant Foam~\cite{govindarajan2025radfoam} & 20k & 26.93 & 0.772 & 0.199 & 23.93 & 30.67 & 137 & 22.17\rfmark{} & 0.788\rfmark{} & 0.184\rfmark{} & 155\rfmark{} \\
2DGS~\cite{Huang2DGS2024} & 30k & 26.81 & 0.796 & 0.214 & 24.23 & 30.04 & 31 & 21.54 & 0.768 & 0.281 & \trd{16} \\
Tetra-NeRF~\cite{kulhanek2023tetranerf} & 300k & 26.40 & 0.727 & 0.269 & 23.36 & 30.21 & 1199 & -- & -- & -- & -- \\
Instant-NGP~\cite{mueller2022instant} & 35k & 25.51 & 0.684 & 0.340 & 22.90 & 28.76 & \fst{4} & 21.62 & 0.712 & 0.340 & \fst{4} \\
\bottomrule
\end{tabular}
\end{table*}

For reference, \cref{tab:supp_others} compares our method with NeRF~\cite{mildenhall2020nerf} and Gaussian splatting~\cite{kerbl3Dgaussians} methods, as evaluated by NerfBaselines~\cite{kulhanek2024nerfbaselines}, and with our baselines.
On Mip-NeRF~360 dataset~\cite{barron2022mipnerf360}, only Zip-NeRF~\cite{barron2023zipnerf} and 3DGS-MCMC~\cite{kheradmand2024mcmc} have a higher PSNR and a lower LPIPS than our method, and Mip-Splatting~\cite{Yu2023MipSplatting} and 3DGS~\cite{kerbl3Dgaussians} have a slightly higher SSIM.
Zip-NeRF trains about $6\times$ longer on Mip-NeRF~360 and $9\times$ longer on Tanks and Temples, where it is ahead of us by 0.15\,dB PSNR and 0.002 SSIM, and we are better in LPIPS.

\subsection{Vertex budget}
\label{sec:supp_budget}

\begin{table}[H]
\centering
\caption{Vertex budget sweep of \cref{fig:budget}, extended with the SSIM, LPIPS and GPU memory of each run. Indoor: Mip-NeRF~360 indoor and Tanks and Temples indoor. RM \#vert.: the per-scene vertex count of Radiance Meshes. GB: peak GPU memory of training, as in \cref{tab:size}. The column min gives the training time in minutes.}
\label{tab:supp_budget}
\footnotesize
\setlength{\tabcolsep}{3pt}
\begin{tabular}{@{}l l ccc rr@{}}
\toprule
Scenes & Budget & PSNR & SSIM & LPIPS & min & GB \\
\midrule
M360 outdoor (5)   & 3M & 24.31 & 0.709 & 0.202 & 37 & 11.9 \\
                   & 4M & 24.39 & 0.715 & 0.193 & 43 & 14.3 \\
                   & 5M & 24.42 & 0.718 & 0.188 & 50 & 17.1 \\
                   & 6M & 24.43 & 0.720 & 0.184 & 56 & 19.9 \\
                   & RM \#vert. & 24.08 & 0.693 & 0.224 & 28 & 9.4 \\
\midrule
T\&T outdoor (15)  & 3M & 25.06 & 0.846 & 0.121 & 37 & 13.6 \\
                   & 4M & 25.04 & 0.847 & 0.119 & 45 & 16.9 \\
                   & 5M & 25.02 & 0.846 & 0.118 & 52 & 20.0 \\
                   & RM \#vert. & 25.04 & 0.843 & 0.135 & 22 & 8.4 \\
\midrule
Indoor (10)        & 2M & 26.72 & 0.862 & 0.127 & 39 & 13.0 \\
                   & 3M & 26.72 & 0.862 & 0.124 & 48 & 15.8 \\
                   & 4M & 26.69 & 0.862 & 0.123 & 57 & 18.6 \\
                   & RM \#vert. & 26.69 & 0.861 & 0.133 & 31 & 10.7 \\
\bottomrule
\end{tabular}
\end{table}

\begin{figure*}[!tp]
\centering
\includegraphics[width=\linewidth]{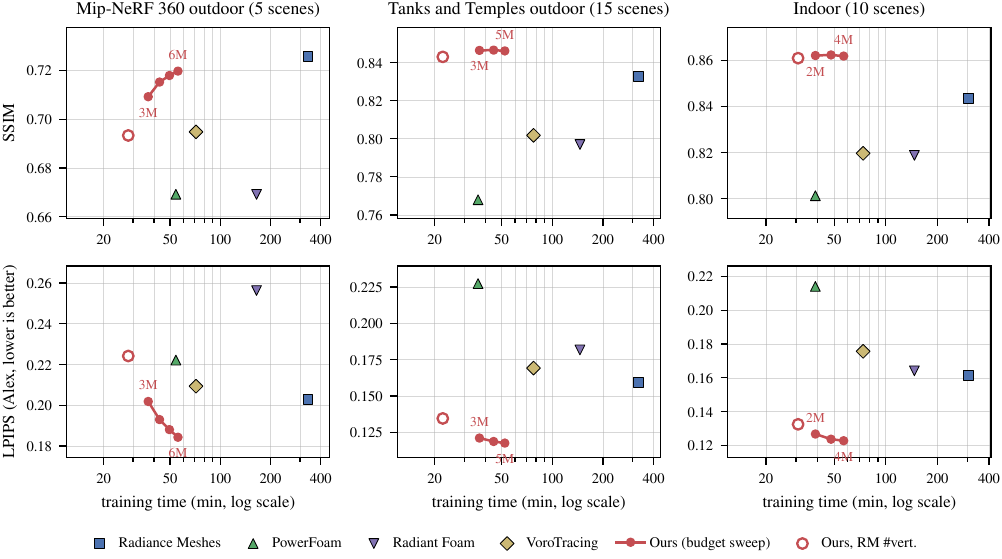}
\caption{SSIM (top) and LPIPS (bottom) against training time (log scale), as in \cref{fig:budget}.}
\label{fig:supp_budget}
\end{figure*}

\Cref{tab:supp_budget} has the numbers of \cref{fig:budget}.
On Mip-NeRF~360 outdoor every metric improves up to 6M.
On Tanks and Temples outdoor and indoors, PSNR and SSIM are flat over all budgets, and only LPIPS improves with more vertices.
Training memory grows linearly with the budget, and every run fits on a 40\,GB A100.
\Cref{fig:supp_budget} shows SSIM and LPIPS against training time, as \cref{fig:budget} shows PSNR.
On Tanks and Temples and indoors, every budget of ours is ahead of every baseline.
On Mip-NeRF~360 outdoor no budget reaches Radiance Meshes in PSNR and SSIM, but from 3M on we are better in LPIPS.

\subsection{Failure cases}
\label{sec:supp_failures}

\begin{figure*}[!tp]
\centering
\setlength{\tabcolsep}{1pt}
\renewcommand{\arraystretch}{0.5}
\begin{tabular}{cccc}
Radiance Meshes & PowerFoam & VoroTracing & Ours \\
\includegraphics[width=0.245\linewidth]{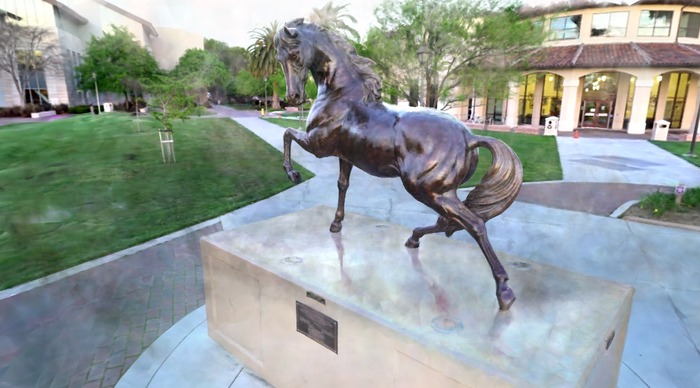} &
\includegraphics[width=0.245\linewidth]{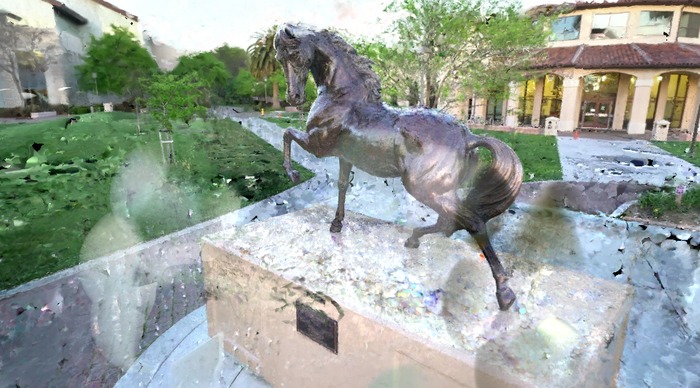} &
\includegraphics[width=0.245\linewidth]{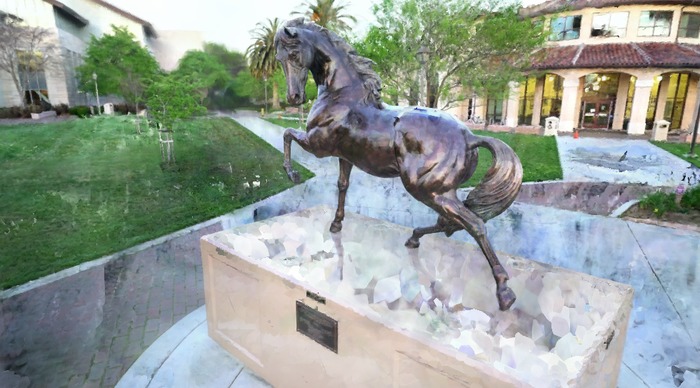} &
\includegraphics[width=0.245\linewidth]{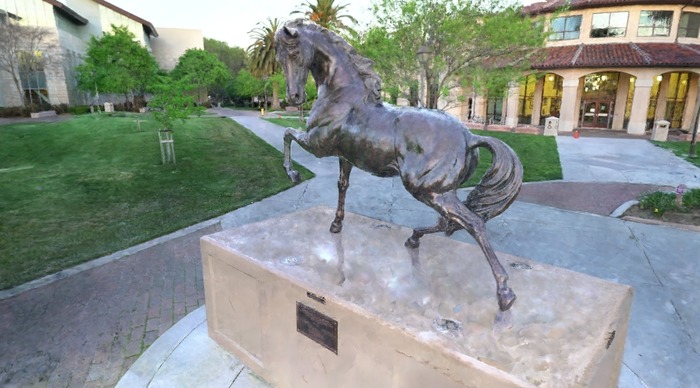} \\
\includegraphics[width=0.245\linewidth]{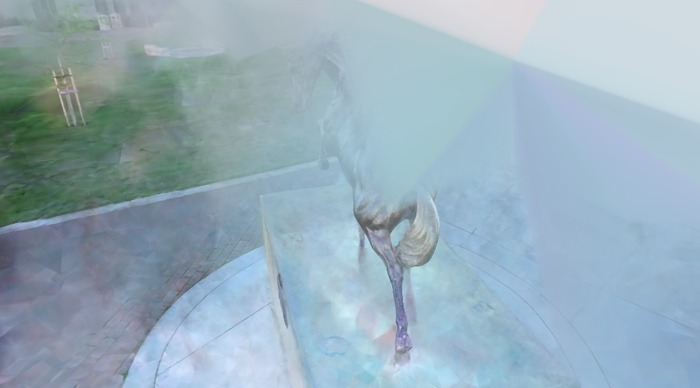} &
\includegraphics[width=0.245\linewidth]{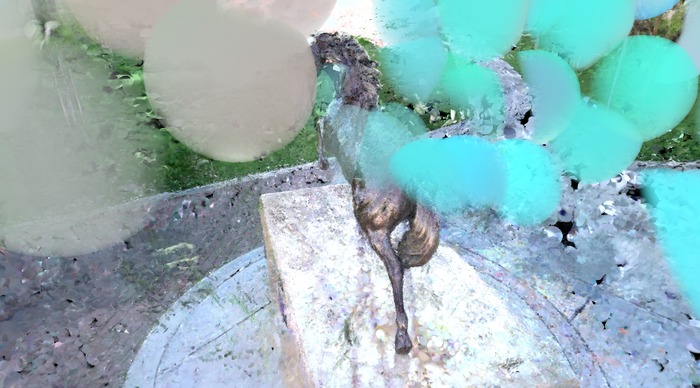} &
\includegraphics[width=0.245\linewidth]{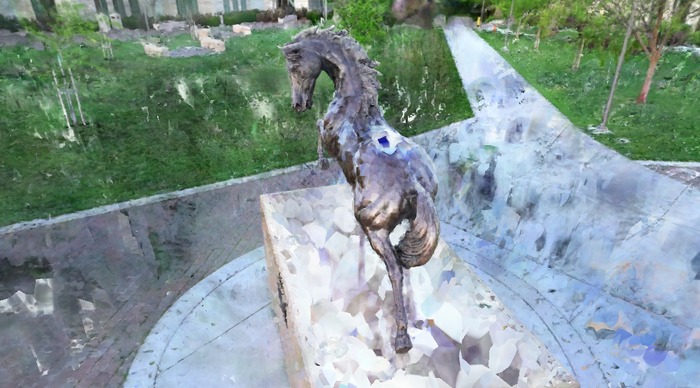} &
\includegraphics[width=0.245\linewidth]{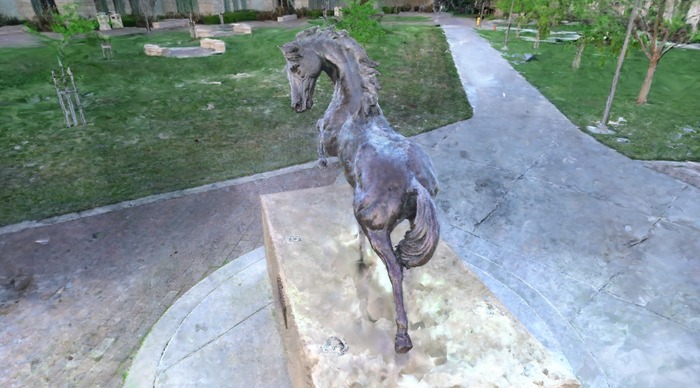} \\
\includegraphics[width=0.245\linewidth]{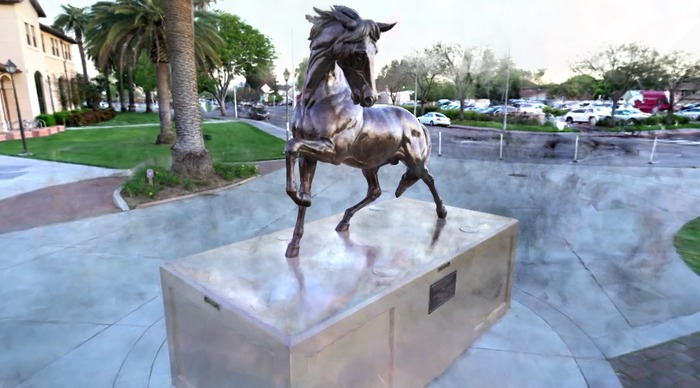} &
\includegraphics[width=0.245\linewidth]{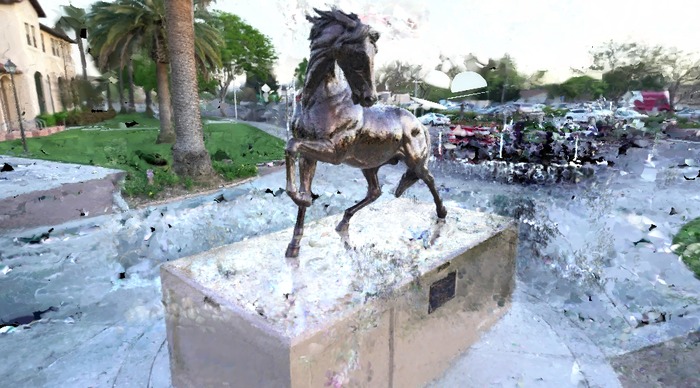} &
\includegraphics[width=0.245\linewidth]{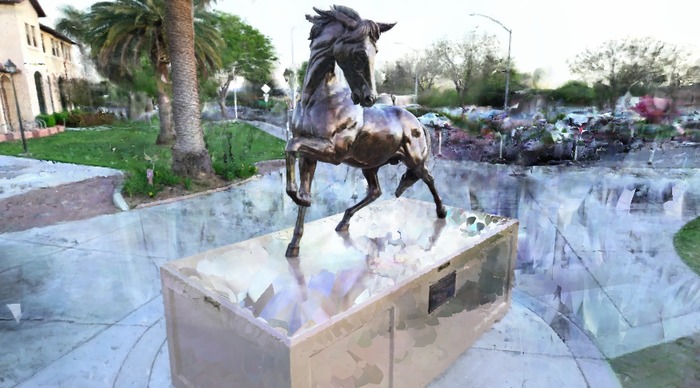} &
\includegraphics[width=0.245\linewidth]{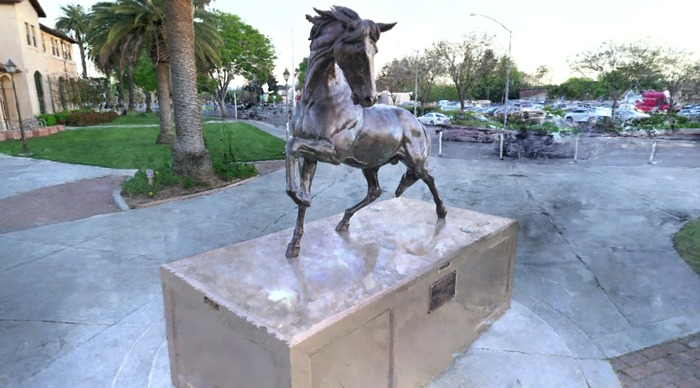} \\
\end{tabular}
\caption{The Horse scene of Tanks and Temples rendered away from the training views, with the models of \cref{tab:nvs}.
The cameras are on the ring of the training cameras, raised by a quarter of its radius (top and bottom rows) or by half of it (middle row).}
\label{fig:supp_floaters}
\end{figure*}

\begin{figure*}[!tp]
\centering
\includegraphics[width=\linewidth]{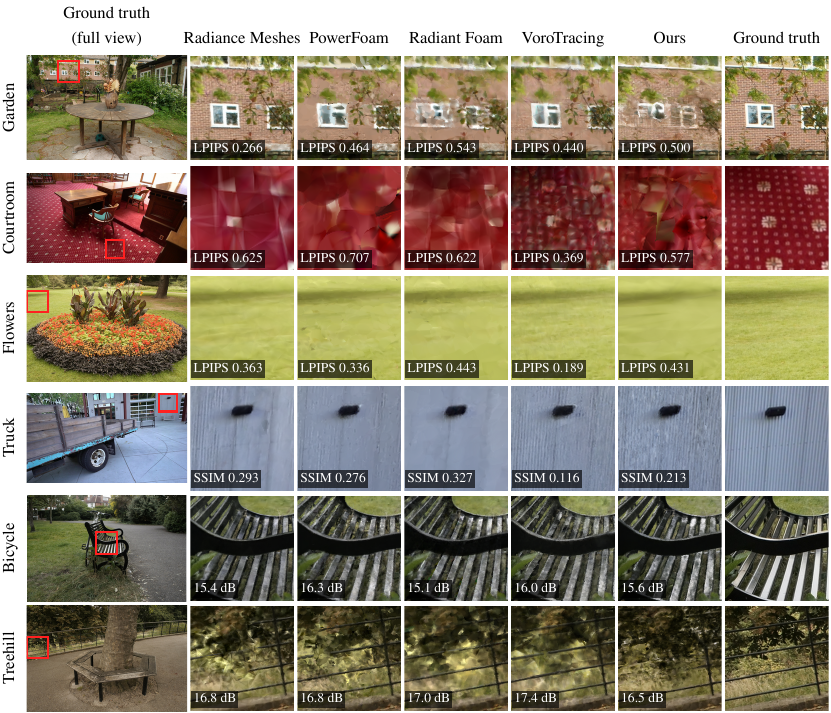}
\caption{Failure cases on test views.
Rows 1--3 show windows where we lose most to the best of the four baselines of \cref{tab:nvs}, rows 4--6 windows where even the best of all methods is worst.
The windows are ranked by local LPIPS (rows 1--3), SSIM (row 4) or PSNR (rows 5 and 6), with that local metric in each crop.
Left, the full ground-truth view with the crop box.}
\label{fig:supp_fail_crops}
\end{figure*}

Away from the training views, all methods show floaters and a broken-up far background (\cref{fig:supp_floaters}).
Seen from above the training cameras of Horse, every method leaves floaters in the air or on the pedestal and breaks up parts of the pavement and the background.

We compare the methods on square windows of each test view, 0.2 times its shorter side.
Our worst windows are fine repeated structure, a patterned carpet and grass close-ups (rows 1--3 of \cref{fig:supp_fail_crops}).
There our mesh breaks into large flat tetrahedra that wipe out texture.
Radiance Meshes is the only method that keeps the window and the brick wall, and VoroTracing keeps more texture than ours in all three.
The corrugated wall of Truck, the slats of the Bicycle bench and the railing in front of the Treehill foliage are hard for all methods (rows 4--6).

\section{Efficiency}
\label{sec:supp_efficiency}

\subsection{Training time}
\label{sec:supp_time}

\begin{figure}[H]
\centering
\includegraphics[width=\columnwidth]{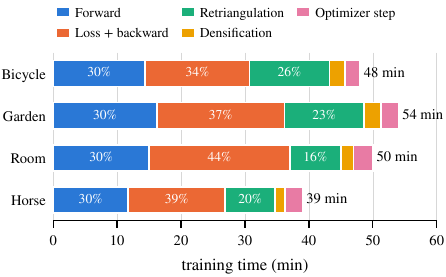}
\caption{Training time per phase on four scenes, without evaluation. Bicycle and garden have 5M vertices, room and horse 3M. Retriangulation is done with gDel3D. Densification takes 4--5\,\% and the optimizer step 5--7\,\% of the training time.}
\label{fig:time}
\end{figure}

Retriangulation takes about a fifth of the training time, 21\,\% on average, and the forward and backward passes take most of the rest, 69\,\% (\cref{fig:time}).

\begin{table}[H]
\centering
\caption{Growing retriangulation interval on the Mip-NeRF~360 outdoor scenes. The interval starts at 10 iterations, grows by 2, 4 or 8 after every retriangulation up to 100, and restarts at 10 after each densification step, as in Radiant Foam~\cite{govindarajan2025radfoam} (which starts at 1 and grows by 2). Retri.: retriangulations before the mesh is fixed at 18k. Rows below ours are differences to it.}
\label{tab:retri}
\footnotesize
\setlength{\tabcolsep}{4pt}
\begin{tabular}{@{}l r ccc r@{}}
\toprule
Interval & Retri. & PSNR$\uparrow$ & SSIM$\uparrow$ & LPIPS$\downarrow$ & Train (min) \\
\midrule
10 (ours)        & 1799 & 24.42 & 0.718 & 0.188 & 49.6 \\
\midrule
10, $+2$ each    & 571  & $-$0.09 & $-$0.003 & $-$0.002 & $-$8.3 \\
10, $+4$ each    & 416  & $-$0.12 & $-$0.005 & $-$0.002 & $-$9.5 \\
10, $+8$ each    & 325  & $-$0.14 & $-$0.005 & $-$0.002 & $-$10.0 \\
\bottomrule
\end{tabular}
\end{table}

\paragraph{Retriangulation schedule.} Radiant Foam~\cite{govindarajan2025radfoam} retriangulates less and less often between densification steps. We tried this schedule starting from our interval of 10 iterations (\cref{tab:retri}). The positions still take one step before each retriangulation, with the gradients accumulated over the interval. It cuts the retriangulations by 3--5$\times$ and the training time by 18--22\%, but PSNR drops by 0.09--0.14~dB on every scene, and more the faster the interval grows. We keep the fixed interval of 10.

\begin{table}[H]
\centering
\caption{Speedups of our implementation. Source: the step of the 3DGS CUDA blog~\cite{kopanas2026cudablog} the change takes, or ours. $^*$Adapted to tetrahedra. Skipping faint segments follows the $1/255$ rule of 3DGS, with our threshold $0.001$. The kernel changes keep the output up to rounding. $^a$Changes training. Single changes are timed on a room checkpoint, the last three rows are full training runs on bicycle, garden and room against our version without the changes. All were measured with an earlier configuration of our method.}
\label{tab:supp_speed}
\scriptsize
\setlength{\tabcolsep}{2.5pt}
\begin{tabular}{@{}l c l@{}}
\toprule
Change & Source & Effect \\
\midrule
Exact tile test & 1.2$^*$ & $-$29\,\% tile pairs, 547 $\to$ 405 entries per pixel \\
Two-stage sort & 1.4 & binning 10 $\to$ 5\,ms \\
Precomputed face planes & ours & forward $-$30\,\% \\
Two-phase forward & ours & forward 23.4 $\to$ 19\,ms \\
Contributor compaction & 4$^*$ & backward $-$6\,\% \\
Cheaper warp reduction & 2$^*$ & backward 76 $\to$ 57\,ms \\
Cheaper position gradients & ours & backward 53 $\to$ 50\,ms \\
Less local memory & 3$^*$ & stack 112 $\to$ 16\,B \\
\midrule
No star splaying$^a$ & ours & retriangulation at 2M vertices 460 $\to$ 230\,ms \\
Skipping faint segments$^a$ & 3DGS & $1.16$--$1.23\times$, 4\,\% with the faint reg. \\
\midrule
All kernel optimizations & & $1.50$--$1.67\times$ training \\
+ no star splaying & & $1.79$--$1.88\times$ training \\
+ skipping faint segments & & $2.08$--$2.33\times$ training, $-$0.00 to $-$0.05\,dB \\
\bottomrule
\end{tabular}
\end{table}

\paragraph{Implementation speedups.} \Cref{tab:supp_speed} lists the speedups of our rasterizer one by one.
From the 3DGS CUDA blog~\cite{kopanas2026cudablog}, we take the two-stage sort of the tile lists and adapt three of its kernel optimizations to tetrahedra, an exact tile test, a backward pass that walks only the contributors recorded by the forward pass, and a cheaper gradient reduction within a warp.
Our backward pass already reduced the gradients within a warp before, so that each warp adds the gradients of a tetrahedron with one atomic operation per value.
Following the blog, the two halves of a warp now take alternate tetrahedra and are combined by one shuffle, which makes this reduction cheaper.
Our own optimizations precompute the face planes of each tetrahedron for the ray tests, let the forward pass first test all tetrahedra of a batch and then composite only the hits, and compute the position gradients with fewer operations.
The color MLP runs in bf16.

\subsection{Rendering speed}
\label{sec:supp_speed}

\begin{table}[H]
\centering
\caption{Rendering speed in FPS, means over the scenes at test resolution on an A100-40GB, measured with \texttt{nerfbaselines measure-fps}.}
\label{tab:supp_fps}
\footnotesize
\setlength{\tabcolsep}{4pt}
\begin{tabular}{@{}l ccc@{}}
\toprule
 & Mip-NeRF~360 (9) & T\&T (21) & All 30 \\
\midrule
Ours        & 29.7 & 45.3 & 40.6 \\
Ours + SS   & 7.6 & 11.7 & 10.4 \\
\bottomrule
\end{tabular}
\end{table}

\Cref{tab:supp_fps} reports our rendering speed.
Supersampling traces 4 rays per pixel and costs about $4\times$ the render time.
We do not compare these speeds with the baselines, since each of them has a dedicated fast renderer besides the renderer used in training, and NerfBaselines measures only the latter.
Measured the same way, the NerfBaselines renderers of the baselines reach 18 to 63\,FPS on average over the 30 scenes, similar to ours.

\paragraph{Interactive viewer.}
\label{sec:supp_viewer}
Our WebGPU viewer renders a trained model in the browser with the same segment integral and color MLP as training, without supersampling.
A trained model is exported once into binary files with the vertex positions, the tetrahedra, the vertex densities and features and the MLP weights, which a static web page loads into GPU memory.
The GPU's hardware rasterizer first draws the boundary faces of the non-empty part of the mesh, which gives each pixel the first non-empty tetrahedron along its ray.
From there, the ray walks from tetrahedron to tetrahedron through their shared faces and composites the features of its segments.
The color MLP then decodes the composited features once per pixel in a compute pass.
The user can orbit, pan and fly through the scene, jump to the test views, and switch between the color, the ground truth, the error, the opacity, the depth and the normals.

\section{Additional ablations and analyses}
\label{sec:supp_ablations}

\subsection{Ablations per scene type}
\label{sec:supp_abl_types}

\begin{table*}[!tp]
\centering
\caption{\Cref{tab:abl} on Mip-NeRF~360, split into outdoor and indoor scenes. It extends \cref{tab:abl} with the rows With priors, With culling at 0.2 and With culling at 0.1, and leaves out its direction-encoding rows, which \cref{tab:supp_scene} splits by scene type. The column min gives the training time in minutes.}
\label{tab:supp_abl_m360}
\footnotesize
\setlength{\tabcolsep}{2.5pt}
\begin{tabular}{@{}l cccc cccc@{}}
\toprule
 & \multicolumn{4}{c}{Mip-NeRF~360 outdoor (5)} & \multicolumn{4}{c}{Mip-NeRF~360 indoor (4)} \\
\cmidrule(lr){2-5}\cmidrule(lr){6-9}
Method & PSNR & SSIM & LPIPS & min & PSNR & SSIM & LPIPS & min \\
\midrule
Ours                     & 24.42 & 0.718 & 0.188 & 50 & 32.24 & 0.930 & 0.106 & 63 \\
\midrule
+ SS                     & +0.20 & +0.006 & +0.011 & +0 & +0.07 & +0.001 & +0.004 & +0 \\
\midrule
With priors              & +0.08 & +0.006 & $-$0.005 & +0 & +0.04 & +0.001 & $-$0.002 & +0 \\
With culling at 0.2      & $-$0.11 & $-$0.002 & +0.002 & $-$4 & $-$0.21 & $-$0.001 & +0.001 & $-$5 \\
With culling at 0.1      & $-$0.08 & $-$0.001 & +0.001 & $-$2 & $-$0.07 & 0.000 & 0.000 & $-$8 \\
Features: hash grid at vertices & $-$0.20 & $-$0.008 & +0.007 & +146 & $-$1.52 & $-$0.011 & +0.017 & +91 \\
No per-tetrahedron densities & $-$0.04 & $-$0.004 & +0.004 & $-$1 & $-$0.08 & $-$0.001 & +0.001 & $-$1 \\
No faint-segment regularizer & +0.02 & +0.002 & +0.001 & +3 & +0.04 & +0.001 & +0.001 & +8 \\
Density: arithmetic mean & $-$0.40 & $-$0.018 & +0.014 & +3 & $-$0.42 & $-$0.002 & +0.001 & +1 \\
Density: minimum         & $-$0.42 & $-$0.026 & +0.029 & +26 & $-$2.74 & $-$0.052 & +0.036 & +10 \\
Color: MLP width 128     & $-$0.08 & $-$0.002 & +0.002 & $-$3 & $-$0.26 & $-$0.002 & +0.002 & $-$4 \\
Color: MLP width 64      & $-$0.14 & $-$0.005 & +0.004 & $-$5 & $-$0.43 & $-$0.003 & +0.004 & $-$5 \\
Color: MLP width 32      & $-$0.29 & $-$0.011 & +0.009 & $-$6 & $-$0.93 & $-$0.007 & +0.008 & $-$5 \\
Color: SH instead of MLP & $-$0.05 & +0.001 & +0.003 & +13 & $-$1.58 & $-$0.010 & +0.016 & +6 \\
Color: SH, colors accumulated & $-$0.12 & $-$0.010 & +0.013 & +46 & $-$2.18 & $-$0.025 & +0.038 & +20 \\
\bottomrule
\end{tabular}
\end{table*}

\begin{table*}[!tp]
\centering
\caption{\Cref{tab:abl} on Tanks and Temples, split into outdoor and indoor scenes. It extends \cref{tab:abl} with the rows With priors, With culling at 0.2 and With culling at 0.1, and leaves out its direction-encoding rows, which \cref{tab:supp_scene} splits by scene type. The column min gives the training time in minutes.}
\label{tab:supp_abl_tnt}
\footnotesize
\setlength{\tabcolsep}{2.5pt}
\begin{tabular}{@{}l cccc cccc@{}}
\toprule
 & \multicolumn{4}{c}{Tanks and Temples outdoor (15)} & \multicolumn{4}{c}{Tanks and Temples indoor (6)} \\
\cmidrule(lr){2-5}\cmidrule(lr){6-9}
Method & PSNR & SSIM & LPIPS & min & PSNR & SSIM & LPIPS & min \\
\midrule
Ours                     & 25.06 & 0.846 & 0.121 & 37 & 23.04 & 0.817 & 0.136 & 38 \\
\midrule
+ SS                     & +0.13 & +0.004 & +0.010 & +0 & +0.08 & +0.004 & +0.007 & +0 \\
\midrule
With priors              & +0.09 & +0.002 & $-$0.001 & +1 & 0.00 & +0.005 & $-$0.007 & $-$1 \\
With culling at 0.2      & $-$0.34 & $-$0.005 & +0.005 & $-$2 & $-$0.17 & $-$0.005 & +0.007 & $-$3 \\
With culling at 0.1      & $-$0.12 & $-$0.002 & +0.002 & $-$2 & $-$0.12 & $-$0.002 & +0.003 & $-$1 \\
Features: hash grid at vertices & $-$0.89 & $-$0.012 & +0.023 & +98 & $-$0.74 & $-$0.017 & +0.029 & +97 \\
No per-tetrahedron densities & $-$0.10 & $-$0.002 & +0.004 & $-$3 & $-$0.14 & $-$0.002 & +0.004 & $-$1 \\
No faint-segment regularizer & +0.06 & +0.002 & 0.000 & +3 & +0.06 & +0.003 & $-$0.001 & +2 \\
Density: arithmetic mean & $-$0.50 & $-$0.011 & +0.016 & +0 & $-$0.35 & $-$0.010 & +0.012 & +1 \\
Density: minimum         & $-$2.42 & $-$0.085 & +0.087 & +12 & $-$0.51 & $-$0.015 & +0.019 & +12 \\
Color: MLP width 128     & $-$0.11 & $-$0.002 & +0.003 & $-$1 & $-$0.10 & $-$0.003 & +0.004 & $-$1 \\
Color: MLP width 64      & $-$0.26 & $-$0.005 & +0.006 & $-$2 & $-$0.23 & $-$0.007 & +0.011 & $-$2 \\
Color: MLP width 32      & $-$0.54 & $-$0.009 & +0.012 & $-$2 & $-$0.45 & $-$0.013 & +0.020 & $-$2 \\
Color: SH instead of MLP & $-$1.08 & $-$0.009 & +0.017 & +10 & $-$1.07 & $-$0.021 & +0.025 & +7 \\
Color: SH, colors accumulated & $-$1.45 & $-$0.022 & +0.037 & +46 & $-$1.49 & $-$0.039 & +0.061 & +27 \\
\bottomrule
\end{tabular}
\end{table*}

\Cref{tab:supp_abl_m360,tab:supp_abl_tnt} split the ablations of \cref{tab:abl}.
The minimum as tetrahedron density fails most on Mip-NeRF~360 indoor and Tanks and Temples outdoor.

\subsection{Tetrahedron density}
\label{sec:supp_density}

The Density rows of \cref{tab:abl} replace the geometric mean of \cref{eq:density} by the arithmetic mean or the minimum of the vertex densities.
With the exp activation, the geometric mean is the arithmetic mean of the log densities, and the minimum is the same before and after the activation, so aggregation both before and after the activation is covered.
Each variant replaces the tetrahedron density everywhere, in rendering, binning and the densification statistics.
As in our method, the vertices of the outer sphere are left out of the mean and the minimum.
With the minimum, a tetrahedron's density and its gradient depend only on its lowest vertex, so each vertex is trained by few of its tetrahedra.
With the arithmetic mean, new vertices start as in our method, at the geometric mean of the densities of the vertices of their tetrahedron.
Starting them at the arithmetic mean instead makes training collapse during densification on every scene, by 5--12\,dB.
A new vertex then gets at least a quarter of the density of the densest vertex of its tetrahedron, so densification plants dense vertices in free space next to surfaces, and their tetrahedra become opaque cells in front of the cameras.
A softplus instead of the exp activation does not help the arithmetic mean ($-$0.07\,dB on bicycle, garden and room).
The geometric mean lets one empty vertex empty a tetrahedron, and it spreads the gradient over all four vertices.
Per-tetrahedron densities are only added once the mesh is fixed at 18k (\cref{sec:supp_hparams}), since a retriangulation replaces tetrahedra, and the No per-tetrahedron densities row of \cref{tab:abl} leaves them out.

\subsection{Faint-segment regularizer}
\label{sec:supp_small}

The faint-segment regularizer (\cref{eq:faint}) pushes faint segments below the skip threshold (\cref{eq:skip}), so that more of them are skipped and a pixel composites fewer segments.
It is not expected to change the number of segments that a ray intersects, only how many of them are skipped.
\Cref{tab:supp_faint} shows this on a subset of three scenes, bicycle, garden and room.
On these scenes, without the regularizer, a pixel composites about twice as many segments, and training takes 7 minutes (14\,\%) longer at almost the same quality.
A larger strength $\lambda_\mathrm{f}$ makes more segments faint enough to be skipped, which shortens training but lowers the quality.

\begin{table}[H]
\centering
\caption{Strength $\lambda_\mathrm{f}$ and threshold $\theta_\mathrm{f}$ of the faint-segment regularizer on bicycle, garden and room.
The upper four rows vary the strength, the lower four the threshold.
Inters.\ and Comp.\ are the numbers of segments per pixel that a ray intersects and composites.
Bold marks our default configuration, with the training time in minutes.
The other rows show differences to it, with the training time in percent.}
\label{tab:supp_faint}
\footnotesize
\setlength{\tabcolsep}{3pt}
\begin{tabular}{@{}cc ccc c cc@{}}
\toprule
 & & & & & & \multicolumn{2}{c}{Segments} \\
\cmidrule(l){7-8}
$\theta_\mathrm{f}$ & $\lambda_\mathrm{f}$ & PSNR & SSIM & LPIPS & Train & Inters. & Comp. \\
\midrule
-- & 0 & +0.03 & +0.002 & +0.001 & +14\,\% & 167.3 & 72.9 \\
0.2 & $1 \cdot 10^{-4}$ & +0.01 & +0.001 & 0.000 & +7\,\% & 163.0 & 51.6 \\
\textbf{0.2} & $\mathbf{3 \cdot 10^{-4}}$ & \textbf{27.96} & \textbf{0.838} & \textbf{0.139} & \textbf{51.6} & \textbf{161.4} & \textbf{35.0} \\
0.2 & $1 \cdot 10^{-3}$ & $-$0.12 & $-$0.006 & +0.003 & $-$10\,\% & 155.2 & 15.5 \\
\midrule
0.1 & $3 \cdot 10^{-4}$ & $-$0.08 & 0.000 & 0.000 & $-$3\,\% & 161.7 & 25.0 \\
\textbf{0.2} & $\mathbf{3 \cdot 10^{-4}}$ & \textbf{27.96} & \textbf{0.838} & \textbf{0.139} & \textbf{51.6} & \textbf{161.4} & \textbf{35.0} \\
0.3 & $3 \cdot 10^{-4}$ & 0.00 & 0.000 & 0.000 & +3\,\% & 162.7 & 41.4 \\
0.4 & $3 \cdot 10^{-4}$ & +0.03 & +0.001 & 0.000 & +4\,\% & 162.3 & 45.6 \\
\bottomrule
\end{tabular}
\end{table}

\subsection{Deferred decoding}
\label{sec:supp_deferred}

We accumulate features along the ray and decode them once per pixel (\cref{sec:rasterization}). With an MLP, decoding per segment would run it once per segment instead, as Hahlbohm \etal~\cite{hahlbohm2026compact} do per Gaussian with a 2$\times$16 MLP.
We test the same choice with SH, where it is cheap.
Here each vertex stores degree-3 SH coefficients instead of features, and there is no MLP.
There are two ways to render them.
The first accumulates the SH coefficients along the ray like our features and evaluates them once per pixel, in the ray direction.
The second evaluates the SH of each vertex in the direction from the camera to that vertex, clamps the resulting color, and accumulates the colors along the ray, as 3DGS does per Gaussian.
Radiance Meshes also accumulates colors, which it evaluates per tetrahedron.
Both start at the same 0.5 grey, the SH offset of 3DGS.
They differ only in where the SH are evaluated and clamped, once per pixel after the accumulation or once per vertex before it.
Without view dependence and clamping, they render the same image.
Accumulating the coefficients is better on every scene type, by 0.37\,dB over the 30 scenes (0.07--0.59\,dB per type), with better SSIM and LPIPS, and trains 29\,min faster (\cref{tab:abl,tab:supp_abl_m360,tab:supp_abl_tnt}).
Both lose to the MLP, 1.0 and 1.4\,dB.
The row Color: SH instead of MLP accumulates the SH coefficients and evaluates them once per pixel, and the row Color: SH, colors accumulated evaluates the SH per vertex and accumulates the colors.
With the better of the two, our method reaches 27.2\,dB on Mip-NeRF~360, on par with the baselines (26.9--27.4\,dB), and 23.4\,dB on Tanks and Temples, still above all of them (22.2--23.2\,dB) (\cref{tab:nvs}).

The higher SH orders need a lower learning rate than our features.
At the learning rate of our features, both variants lose 3--6\,dB.
On bicycle, garden and room, we tried this learning rate scaled by 1, 1/20 and 1/100, and the learning rates of 3DGS.
Scaling by 1/20, which is also the ratio between the learning rates of the higher SH orders and of the base color in 3DGS, worked best, so we use it.

\subsection{Densification and culling}
\label{sec:supp_dens}

\Cref{tab:dens} compares where the vertices go, at the vertex count of Radiance Meshes on three scenes.
Its densifier uses error thresholds tuned for its model which do not work with our model, so we replay the insertions recorded in its runs.
Our densification and the replayed insertions both reach the PSNR of Radiance Meshes itself, and ours gives a better SSIM and LPIPS than the replay.
Radiance Meshes keeps a better SSIM and LPIPS at this count, but trains $10\times$ longer.
Our densifier needs no tuned error or density thresholds (\cref{sec:supp_method}).
Our full budget gains over the matched count, and our fast training makes this budget affordable.

\begin{table*}[!tp]
\centering
\caption{Densification at the final vertex count of the recorded Radiance Meshes runs (bicycle 2.0M, garden 1.2M, room 0.82M). Own densification: our selection and schedule up to that count. RM's insertions: the insertion points recorded in Radiance Meshes runs without culling, replayed at their iterations and snapped to our cells, with our cell init. Grey: ours at its own budget. The column train gives the training time in minutes.}
\label{tab:dens}
\footnotesize
\setlength{\tabcolsep}{2.5pt}
\begin{tabular}{@{}l ccc ccc ccc ccc c@{}}
\toprule
 & \multicolumn{3}{c}{bicycle} & \multicolumn{3}{c}{garden} & \multicolumn{3}{c}{room} & \multicolumn{3}{c}{mean} & train \\
\cmidrule(lr){2-4}\cmidrule(lr){5-7}\cmidrule(lr){8-10}\cmidrule(lr){11-13}
Method & PSNR & SSIM & LPIPS & PSNR & SSIM & LPIPS & PSNR & SSIM & LPIPS & PSNR & SSIM & LPIPS & min \\
\midrule
Radiance Meshes & 24.98 & 0.740 & 0.208 & 26.99 & 0.849 & 0.105 & 30.58 & 0.913 & 0.167 & 27.51 & 0.834 & 0.160 & 296 \\
Ours, own densification & 24.32 & 0.710 & 0.216 & 26.46 & 0.816 & 0.136 & 31.84 & 0.922 & 0.147 & 27.54 & 0.816 & 0.167 & 29 \\
Ours, RM's insertions & 24.50 & 0.690 & 0.262 & 26.70 & 0.815 & 0.139 & 31.52 & 0.912 & 0.167 & 27.57 & 0.806 & 0.189 & 29 \\
\midrule
\old{Ours (5M / 3M)} & \old{24.74} & \old{0.741} & \old{0.182} & \old{27.02} & \old{0.847} & \old{0.102} & \old{32.11} & \old{0.926} & \old{0.133} & \old{27.96} & \old{0.838} & \old{0.139} & \old{52} \\
\bottomrule
\end{tabular}
\end{table*}

\begin{figure}[H]
\centering
\setlength{\tabcolsep}{1pt}
\renewcommand{\arraystretch}{0.5}
\begin{tabular}{cc}
Ours & With culling at 0.2 \\
\includegraphics[width=0.49\linewidth]{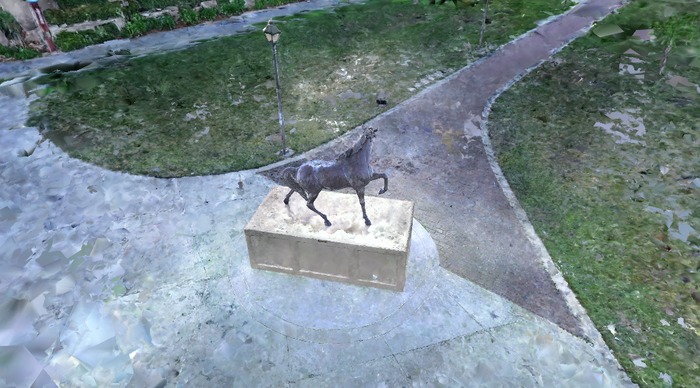} &
\includegraphics[width=0.49\linewidth]{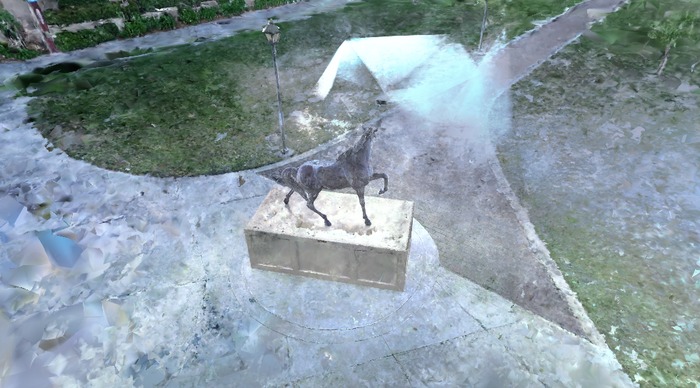} \\
\includegraphics[width=0.49\linewidth]{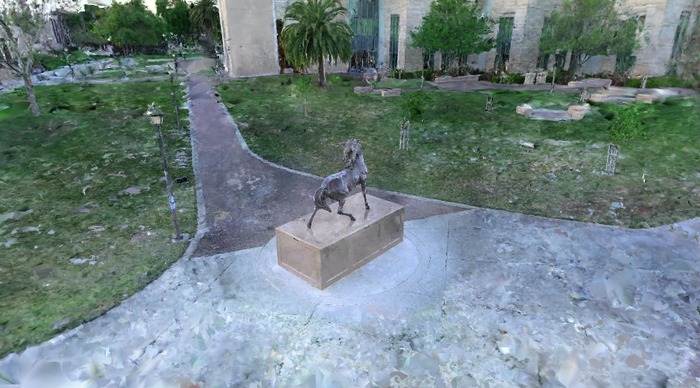} &
\includegraphics[width=0.49\linewidth]{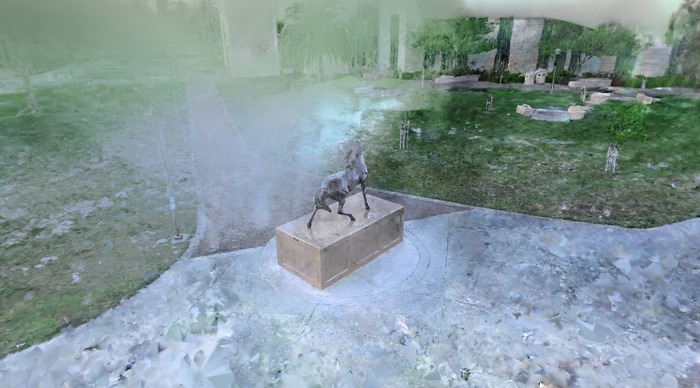} \\
\includegraphics[width=0.49\linewidth]{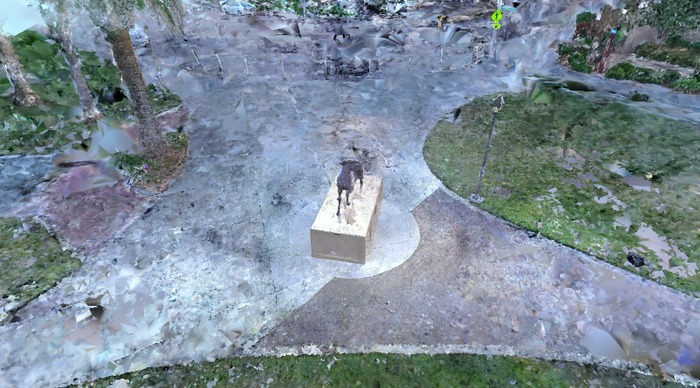} &
\includegraphics[width=0.49\linewidth]{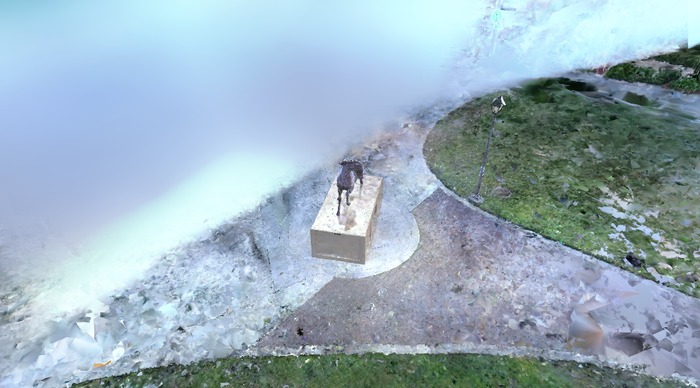} \\
\end{tabular}
\caption{The Horse scene of Tanks and Temples seen from far above the training cameras, at two (top row) and three times (middle and bottom rows) the radius of their ring.
With culling, large white floaters hang over the plaza, while our model without culling stays clear.}
\label{fig:supp_cull_floaters}
\end{figure}

\paragraph{Culling.} We expected culling (\cref{eq:cull}) to remove floaters, but the culled models actually show more of them (\cref{fig:supp_cull_floaters}).
Culling also costs PSNR, most on Tanks and Temples, and shortens our training by only 3\,min (\cref{tab:supp_abl_m360,tab:supp_abl_tnt}).
With culling at 0.1, the threshold of Radiance Meshes, costs about half as much over the 30 scenes.
Without it, Radiance Meshes is 0.16\,dB better over the 30 scenes, but trains 386 instead of 320\,min.
Our method is better than Radiance Meshes without culling on 26 of the 30 scenes in PSNR, on 27 in SSIM and on all 30 in LPIPS.
So for Radiance Meshes culling mainly saves time, and our training is fast without it.
Over the 30 scenes, culling costs 0.25\,dB, and culling at 0.1 costs 0.10\,dB.

\subsection{Retriangulation probe}
\label{sec:supp_probe}

\begin{table}[H]
\centering
\caption{Retriangulation probe on bicycle, garden and room, means per phase of training.
New tets is the share of tetrahedra that a retriangulation replaces, and Verts the share of vertices these tetrahedra touch.
PSNR and Pixels compare the renders before and after the flips, and before and after the position step.
$\Delta$GT is the change of PSNR against the ground truth caused by the flips, in dB.}
\scriptsize
\setlength{\tabcolsep}{2.5pt}
\begin{tabular}{@{}l l ccccccc@{}}
\toprule
 & & & & \multicolumn{2}{c}{Flips} & \multicolumn{2}{c}{Position step} & \\
\cmidrule(lr){5-6}\cmidrule(lr){7-8}
Scene & Iterations & New tets & Verts & PSNR & Pixels & PSNR & Pixels & $\Delta$GT \\
\midrule
bicycle & 0--5k   & 9.3\,\% & 45\,\% & 41.6 & 5.9\,\% & 48.5 & 2.0\,\% & $-$0.016 \\
        & 5k--10k & 4.2\,\% & 31\,\% & 45.4 & 3.1\,\% & 51.9 & 1.3\,\% & $-$0.016 \\
        & 10k--18k & 1.6\,\% & 13\,\% & 48.4 & 1.3\,\% & 58.5 & 0.5\,\% & $-$0.009 \\
garden  & 0--5k   & 7.8\,\% & 47\,\% & 41.7 & 8.6\,\% & 48.4 & 2.4\,\% & $-$0.042 \\
        & 5k--10k & 3.8\,\% & 31\,\% & 48.6 & 2.7\,\% & 54.1 & 0.9\,\% & $-$0.014 \\
        & 10k--18k & 1.5\,\% & 14\,\% & 51.4 & 1.2\,\% & 58.9 & 0.3\,\% & $-$0.008 \\
room    & 0--5k   & 6.2\,\% & 43\,\% & 50.0 & 2.0\,\% & 57.8 & 0.4\,\% & $-$0.019 \\
        & 5k--10k & 3.0\,\% & 27\,\% & 56.2 & 0.5\,\% & 62.8 & 0.1\,\% & $-$0.087 \\
        & 10k--18k & 1.2\,\% & 12\,\% & 56.4 & 0.3\,\% & 66.1 & 0.04\,\% & $-$0.051 \\
\bottomrule
\end{tabular}
\label{tab:supp_probe}
\end{table}

\begin{table}[H]
\centering
\caption{Mesh changes and training loss at the retriangulations, means per phase of training.
Tets is the number of tetrahedra before a retriangulation, and Created and Removed are the tetrahedra it creates and removes.
Loss is the change of the training loss right after a retriangulation, and SE its standard error, both in percent.}
\scriptsize
\setlength{\tabcolsep}{3pt}
\begin{tabular}{@{}l l ccc rr@{}}
\toprule
Scene & Iterations & Tets & Created & Removed & Loss & SE \\
\midrule
bicycle & 0--5k    & 6.7M  & 624k (9.3\,\%) & 623k & +0.26\,\% & 0.42\,\% \\
        & 5k--10k  & 30.4M & 1282k (4.2\,\%) & 1282k & +0.27\,\% & 0.18\,\% \\
        & 10k--18k & 32.6M & 519k (1.6\,\%) & 519k & +0.05\,\% & 0.08\,\% \\
garden  & 0--5k    & 8.2M  & 645k (7.8\,\%) & 644k & +0.10\,\% & 0.52\,\% \\
        & 5k--10k  & 30.8M & 1163k (3.8\,\%) & 1163k & +0.14\,\% & 0.15\,\% \\
        & 10k--18k & 32.6M & 492k (1.5\,\%) & 492k & +0.06\,\% & 0.06\,\% \\
room    & 0--5k    & 5.4M  & 333k (6.2\,\%) & 333k & +0.81\,\% & 0.92\,\% \\
        & 5k--10k  & 18.6M & 552k (3.0\,\%) & 552k & +0.10\,\% & 0.29\,\% \\
        & 10k--18k & 19.6M & 241k (1.2\,\%) & 241k & $-$0.07\,\% & 0.13\,\% \\
\bottomrule
\end{tabular}
\label{tab:supp_probe_loss}
\end{table}

Every 10 iterations, the vertices take a position step and the mesh is retriangulated.
The retriangulation flips edges, \ie it replaces some tetrahedra by others over the same vertices.
To see what these flips do, we probe every 5th retriangulation of a full training run on bicycle, garden and room.
We average the probes over three phases of training, before 5k iterations, from 5k to 10k, and from 10k until the mesh is fixed at 18k.

\paragraph{Mesh changes.} We compare the tetrahedra before and after a retriangulation, each identified by its four vertices.
A retriangulation replaces 6--9\,\% of the tetrahedra by new ones before 5k iterations, 3--4\,\% until 10k and 1.2--1.6\,\% until 18k (\cref{tab:supp_probe_loss}).
It removes about as many tetrahedra as it creates and leaves all others unchanged.
The new tetrahedra touch 12--47\,\% of the vertices (\cref{tab:supp_probe}).

\paragraph{Renders.} At each probe, we render 4 fixed training views three times, before the position step, after it but still with the old tetrahedra, and after the retriangulation.
We compare the render before the position step with the one after it, which shows the effect of the position step alone.
We also compare the render after the position step with the one after the retriangulation, which shows the effect of the flips alone.
For each of these two pairs, we measure the PSNR between the two renders and the share of pixels that change by more than 2/255 (\cref{tab:supp_probe}).
The flips change the renders more than the position step, but only in 0.3--9\,\% of the pixels.
Against the ground truth, they lower PSNR by 0.01--0.09\,dB.

\paragraph{Training loss.} We also log the training loss of every iteration.
To remove the difficulty of each view and the training trend, we divide each loss by the geometric mean of the losses of the same view within the same 1000 iterations.
We then compare the first iteration after a retriangulation with the other 9 iterations until the next retriangulation, over all retriangulations of a phase, and leave out the 100 iterations after each densification step.
The loss after a retriangulation is at most 0.8\,\% higher, within 1.5 standard errors of the mean (\cref{tab:supp_probe_loss}), and the other 9 iterations vary by as much, so the flips do not measurably disturb training.

\paragraph{Topology freeze and position learning rate.} On bicycle, garden and room, never fixing the mesh, with position steps and retriangulations until the end of training and hence without per-tetrahedron densities, changes PSNR by $-0.09$ to $+0.04$\,dB per scene, $-0.01$\,dB on average, and trains 11\,min longer.
Fixing the mesh at 18k thus saves time but is not needed for quality.
Twice the position learning rate costs 0.04\,dB on average, and four times costs 0.25\,dB, most on bicycle ($-0.71$\,dB).

\subsection{Monocular priors}
\label{sec:supp_depth}

The With priors row of \cref{tab:supp_abl_m360,tab:supp_abl_tnt} adds two monocular priors that our method does not use by default.
The first is a normal loss in the first 5k iterations.
For each pixel, we composite the normals of the faces through which the ray enters the tetrahedra, with the blending weights.
The loss is $\lambda_\mathrm{n} (1 - \cos\theta_p)$, averaged over the pixels with accumulated opacity above 0.5, where $\theta_p$ is the angle between this normal and the MoGe-2~\cite{wang2025moge2} normal of the pixel, and $\lambda_\mathrm{n} = 0.01$.
MeshSplatting~\cite{Held2025MeshSplatting} uses the same loss.
The second masks the densification with the depth of Depth Anything 3~\cite{depthanything3} instead of the rendered depth.
We tuned the hyperparameters of both priors, and together they slightly improve our method.
We nevertheless dropped them for simplicity and to stay comparable with the baselines, none of which uses monocular priors.

Together they gain only 0.06\,dB on Mip-NeRF~360 and 0.07\,dB on Tanks and Temples (With priors, \cref{tab:supp_abl_m360,tab:supp_abl_tnt}).
A better depth source does not help either.
Multi-view depth from VGGT-Omega~\cite{wang2026vggtomega} masks the densification no better than DA3 (\cref{tab:supp_depth}).

\begin{table}[H]
\centering
\caption{Depth sources. All rows use culling, the MoGe-2 normal loss and per-vertex densities, unlike our main results, so they are lower than ours in \cref{tab:nvs}. The first row is the rendered-depth mask of our method in this setting, and the other rows are differences to it. Upper rows: the source only masks the densification. Lower rows: it also drives the depth loss (weight 0.1 until 5k). DA3 aligned: DA3 scaled per view to the SfM points. VGGT-Omega: multi-view depth. --: not run.}
\label{tab:supp_depth}
\scriptsize
\setlength{\tabcolsep}{2.5pt}
\begin{tabular}{@{}l ccc ccc@{}}
\toprule
 & \multicolumn{3}{c}{Mip-NeRF~360 (9)} & \multicolumn{3}{c}{T\&T (21)} \\
\cmidrule(lr){2-4}\cmidrule(lr){5-7}
Depth source & PSNR & SSIM & LPIPS & PSNR & SSIM & LPIPS \\
\midrule
Rendered depth (ours) & 27.66 & 0.810 & 0.154 & 24.05 & 0.831 & 0.133 \\
\midrule
DA3 & +0.03 & +0.001 & $-$0.001 & +0.14 & +0.004 & $-$0.004 \\
DA3 aligned & +0.01 & +0.001 & 0.000 & -- & -- & -- \\
VGGT-Omega & +0.05 & +0.002 & $-$0.001 & +0.13 & +0.004 & $-$0.004 \\
\midrule
DA3 + depth loss & +0.04 & +0.002 & $-$0.002 & $-$0.08 & +0.002 & $-$0.004 \\
DA3 aligned + depth loss & +0.06 & +0.002 & $-$0.001 & -- & -- & -- \\
VGGT-Omega + depth loss & +0.10 & +0.003 & $-$0.002 & $-$0.04 & +0.004 & $-$0.005 \\
\bottomrule
\end{tabular}
\end{table}

\Cref{tab:supp_depth} compares the depth sources in a configuration with culling, the MoGe-2 normal loss and per-vertex densities.
In this setting, the DA3 mask is 0.03\,dB better than our rendered-depth mask on Mip-NeRF~360 and 0.14\,dB better on Tanks and Temples.
DA3 predicts each view on its own, so its scale varies between views.
On bicycle, flowers, stump and bonsai many views are off by more than 15\% against the SfM points.
DA3 aligned and VGGT-Omega fix this, but as densification masks they perform about the same as DA3.
A depth loss on top of the mask gains at most 0.05\,dB more on Mip-NeRF~360, with VGGT-Omega, and it helps or hurts the same scenes whichever source drives it.
On Tanks and Temples both DA3 and VGGT-Omega have the right scale in almost every view, and the loss costs about 0.2\,dB PSNR with either source.

\clearpage
\begin{table*}[p]
\centering
\caption{Results on every scene of Mip-NeRF~360, with the runs of \cref{tab:nvs}. RM: Radiance Meshes, PF: PowerFoam, RF: Radiant Foam, VT: VoroTracing, Ours: our method without supersampling, +SS: ours with supersampling. Bold: the best of the first five methods. The last row is the mean of \cref{tab:nvs}.}
\label{tab:supp_scenes_m360}
\scriptsize
\setlength{\tabcolsep}{2.2pt}
\begin{tabular}{@{}l cccccc cccccc cccccc@{}}
\toprule
 & \multicolumn{6}{c}{PSNR$\uparrow$} & \multicolumn{6}{c}{SSIM$\uparrow$} & \multicolumn{6}{c}{LPIPS$\downarrow$} \\
\cmidrule(lr){2-7}\cmidrule(lr){8-13}\cmidrule(lr){14-19}
Scene & RM & PF & RF & VT & Ours & +SS & RM & PF & RF & VT & Ours & +SS & RM & PF & RF & VT & Ours & +SS \\
\midrule
Bicycle      & \textbf{24.98} & 24.07 & 24.14 & 24.34 & 24.74 & 24.98 & 0.740 & 0.675 & 0.672 & 0.709 & \textbf{0.741} & 0.747 & 0.208 & 0.226 & 0.260 & 0.205 & \textbf{0.182} & 0.194 \\
Flowers      & \textbf{21.61} & 21.31 & 21.35 & 21.54 & 21.41 & 21.64 & \textbf{0.626} & 0.573 & 0.568 & 0.589 & 0.608 & 0.616 & 0.250 & 0.271 & 0.312 & 0.269 & \textbf{0.245} & 0.257 \\
Garden       & 26.99 & 26.96 & 26.56 & 27.00 & \textbf{27.02} & 27.16 & \textbf{0.849} & 0.824 & 0.809 & 0.822 & 0.847 & 0.848 & 0.105 & \textbf{0.099} & 0.125 & 0.112 & 0.102 & 0.121 \\
Stump        & 26.39 & 25.11 & 25.33 & 25.95 & \textbf{26.50} & 26.72 & 0.761 & 0.689 & 0.706 & 0.738 & \textbf{0.769} & 0.777 & 0.167 & 0.200 & 0.202 & 0.167 & \textbf{0.143} & 0.144 \\
Treehill     & \textbf{22.97} & 22.36 & 22.26 & 22.59 & 22.43 & 22.60 & \textbf{0.652} & 0.586 & 0.591 & 0.616 & 0.624 & 0.630 & 0.284 & 0.316 & 0.382 & 0.294 & \textbf{0.268} & 0.282 \\
Room         & 30.58 & 31.08 & 30.75 & 31.32 & \textbf{32.11} & 32.14 & 0.913 & 0.905 & 0.899 & 0.915 & \textbf{0.926} & 0.927 & 0.167 & 0.160 & 0.160 & 0.139 & \textbf{0.133} & 0.138 \\
Counter      & 28.56 & 29.67 & 28.53 & 29.43 & \textbf{30.14} & 30.22 & 0.905 & 0.894 & 0.874 & 0.893 & \textbf{0.913} & 0.915 & 0.137 & 0.128 & 0.143 & 0.122 & \textbf{0.115} & 0.118 \\
Kitchen      & 30.84 & 31.09 & 31.22 & 31.54 & \textbf{32.11} & 32.15 & 0.922 & 0.914 & 0.905 & 0.919 & \textbf{0.930} & 0.931 & 0.093 & 0.094 & 0.100 & 0.089 & \textbf{0.085} & 0.089 \\
Bonsai       & 30.62 & 33.16 & 32.20 & 33.13 & \textbf{34.60} & 34.72 & 0.933 & 0.936 & 0.924 & 0.934 & \textbf{0.952} & 0.953 & 0.123 & 0.104 & 0.108 & 0.096 & \textbf{0.091} & 0.095 \\
\midrule
Mean         & 27.06 & 27.20 & 26.93 & 27.43 & \textbf{27.90} & 28.04 & 0.811 & 0.777 & 0.772 & 0.793 & \textbf{0.812} & 0.816 & 0.171 & 0.178 & 0.199 & 0.166 & \textbf{0.152} & 0.160 \\
\bottomrule
\end{tabular}
\end{table*}

\begin{table*}[p]
\centering
\caption{Results on every scene of Tanks and Temples, with the runs of \cref{tab:nvs}. RM: Radiance Meshes, PF: PowerFoam, RF: Radiant Foam, VT: VoroTracing, Ours: our method without supersampling, +SS: ours with supersampling. Bold: the best of the first five methods. The last row is the mean of \cref{tab:nvs}. \rfmark{}Radiant Foam uses gDel3D for the triangulation on Train, as its own Delaunay implementation crashes there (\cref{tab:nvs}).}
\label{tab:supp_scenes_tnt}
\scriptsize
\setlength{\tabcolsep}{2.2pt}
\begin{tabular}{@{}l cccccc cccccc cccccc@{}}
\toprule
 & \multicolumn{6}{c}{PSNR$\uparrow$} & \multicolumn{6}{c}{SSIM$\uparrow$} & \multicolumn{6}{c}{LPIPS$\downarrow$} \\
\cmidrule(lr){2-7}\cmidrule(lr){8-13}\cmidrule(lr){14-19}
Scene & RM & PF & RF & VT & Ours & +SS & RM & PF & RF & VT & Ours & +SS & RM & PF & RF & VT & Ours & +SS \\
\midrule
Auditorium   & 22.28 & 21.77 & 20.90 & 21.08 & \textbf{23.57} & 23.61 & 0.841 & 0.759 & 0.812 & 0.776 & \textbf{0.855} & 0.858 & 0.224 & 0.377 & 0.222 & 0.320 & \textbf{0.174} & 0.178 \\
Ballroom     & 23.12 & 23.35 & 22.89 & 23.65 & \textbf{24.55} & 24.70 & 0.798 & 0.767 & 0.769 & 0.789 & \textbf{0.831} & 0.835 & 0.136 & 0.165 & 0.152 & 0.138 & \textbf{0.098} & 0.107 \\
Church       & 21.53 & 21.07 & 19.91 & 21.36 & \textbf{22.88} & 22.94 & 0.791 & 0.707 & 0.761 & 0.766 & \textbf{0.821} & 0.824 & 0.178 & 0.275 & 0.180 & 0.189 & \textbf{0.134} & 0.144 \\
Courtroom    & 20.84 & 20.53 & 19.78 & 20.75 & \textbf{22.88} & 22.97 & 0.758 & 0.681 & 0.723 & 0.722 & \textbf{0.793} & 0.797 & 0.200 & 0.314 & 0.205 & 0.221 & \textbf{0.135} & 0.141 \\
Meetingroom  & 23.68 & 23.79 & 22.69 & 23.91 & \textbf{25.19} & 25.30 & 0.850 & 0.805 & 0.815 & 0.816 & \textbf{0.868} & 0.871 & 0.156 & 0.229 & 0.163 & 0.185 & \textbf{0.113} & 0.119 \\
Museum       & 18.67 & 18.64 & 18.63 & 18.00 & \textbf{19.16} & 19.22 & 0.726 & 0.646 & 0.705 & 0.666 & \textbf{0.734} & 0.739 & 0.198 & 0.295 & 0.210 & 0.257 & \textbf{0.160} & 0.166 \\
Barn         & 27.36 & 27.03 & 26.67 & 27.18 & \textbf{28.38} & 28.55 & 0.865 & 0.827 & 0.832 & 0.843 & \textbf{0.877} & 0.879 & 0.125 & 0.164 & 0.154 & 0.127 & \textbf{0.092} & 0.103 \\
Caterpillar  & 22.82 & 22.31 & 21.98 & 22.93 & \textbf{24.94} & 25.09 & 0.786 & 0.702 & 0.737 & 0.739 & \textbf{0.822} & 0.825 & 0.190 & 0.252 & 0.203 & 0.204 & \textbf{0.137} & 0.152 \\
Courthouse   & 22.12 & 21.98 & 21.30 & 22.65 & \textbf{23.48} & 23.55 & 0.781 & 0.721 & 0.761 & 0.770 & \textbf{0.802} & 0.804 & 0.220 & 0.319 & 0.233 & 0.211 & \textbf{0.179} & 0.187 \\
Family       & 23.75 & 23.22 & 23.21 & 23.38 & \textbf{25.14} & 25.33 & 0.856 & 0.781 & 0.814 & 0.809 & \textbf{0.862} & 0.867 & 0.120 & 0.169 & 0.145 & 0.139 & \textbf{0.079} & 0.090 \\
Francis      & 26.72 & 24.81 & 24.36 & 26.56 & \textbf{30.79} & 30.89 & 0.891 & 0.810 & 0.867 & 0.877 & \textbf{0.921} & 0.923 & 0.165 & 0.289 & 0.168 & 0.155 & \textbf{0.109} & 0.119 \\
Horse        & 23.12 & 22.17 & 22.73 & 23.14 & \textbf{25.38} & 25.59 & 0.869 & 0.780 & 0.827 & 0.833 & \textbf{0.878} & 0.883 & 0.124 & 0.212 & 0.153 & 0.134 & \textbf{0.075} & 0.080 \\
Ignatius     & 20.92 & 20.54 & 20.69 & 20.97 & \textbf{22.00} & 22.13 & 0.757 & 0.667 & 0.716 & 0.708 & \textbf{0.776} & 0.781 & 0.192 & 0.225 & 0.193 & 0.195 & \textbf{0.140} & 0.159 \\
Lighthouse   & 22.70 & 21.87 & 20.39 & 22.09 & \textbf{24.05} & 24.14 & 0.846 & 0.785 & 0.802 & 0.821 & \textbf{0.859} & 0.863 & 0.150 & 0.255 & 0.199 & 0.168 & \textbf{0.113} & 0.122 \\
M60          & 27.91 & 27.19 & 27.42 & 27.50 & \textbf{28.80} & 28.95 & 0.905 & 0.881 & 0.880 & 0.883 & \textbf{0.914} & 0.916 & 0.093 & 0.115 & 0.107 & 0.100 & \textbf{0.073} & 0.080 \\
Palace       & 20.05 & 19.43 & 18.38 & 19.26 & \textbf{20.46} & 20.52 & 0.740 & 0.674 & 0.710 & 0.706 & \textbf{0.750} & 0.753 & 0.282 & 0.390 & 0.324 & 0.326 & \textbf{0.242} & 0.250 \\
Panther      & 28.48 & 28.40 & 28.15 & 28.18 & \textbf{28.75} & 28.88 & 0.906 & 0.892 & 0.887 & 0.887 & \textbf{0.911} & 0.913 & 0.098 & 0.109 & 0.112 & 0.106 & \textbf{0.083} & 0.090 \\
Playground   & 25.58 & 25.24 & 24.16 & 26.05 & \textbf{27.54} & 27.68 & 0.851 & 0.794 & 0.817 & 0.833 & \textbf{0.884} & 0.887 & 0.153 & 0.177 & 0.159 & 0.127 & \textbf{0.094} & 0.112 \\
Temple       & 19.88 & 19.31 & 18.05 & 19.29 & \textbf{20.76} & 20.84 & 0.785 & 0.699 & 0.747 & 0.750 & \textbf{0.793} & 0.797 & 0.223 & 0.351 & 0.257 & 0.257 & \textbf{0.182} & 0.186 \\
Train        & 21.19 & 20.34 & 19.95\rfmark{} & 20.86 & \textbf{21.95} & 22.07 & 0.801 & 0.707 & 0.744 & 0.752 & \textbf{0.809} & 0.814 & 0.146 & 0.241 & 0.182 & 0.168 & \textbf{0.120} & 0.131 \\
Truck        & \textbf{23.73} & 23.28 & 23.26 & 23.40 & 23.46 & 23.63 & \textbf{0.854} & 0.799 & 0.813 & 0.816 & 0.839 & 0.843 & 0.105 & 0.143 & 0.135 & 0.121 & \textbf{0.099} & 0.108 \\
\midrule
Mean         & 23.16 & 22.68 & 22.17 & 22.96 & \textbf{24.48} & 24.60 & 0.822 & 0.756 & 0.788 & 0.789 & \textbf{0.838} & 0.842 & 0.166 & 0.241 & 0.184 & 0.183 & \textbf{0.125} & 0.135 \\
\bottomrule
\end{tabular}
\end{table*}

\end{document}